\documentclass[12pt]{elsarticle}
\usepackage[latin9]{inputenc}
\usepackage{geometry}
\usepackage{color}
\usepackage{amsmath}
\usepackage{amssymb}
\usepackage{cancel}
\usepackage{graphicx}

\makeatletter

\DeclareTextSymbolDefault{\textquotedbl}{T1}
\providecommand{\tabularnewline}{\\}

\usepackage{outlines}
\usepackage{subcaption}
\usepackage{xcolor}
\usepackage{todonotes}
\journal{Journal of Computational Physics}
\DeclareMathOperator\erfc{erfc}

\makeatother

\begin{document}

\title{A fully implicit, asymptotic-preserving, semi-Lagrangian algorithm
for the time dependent anisotropic heat transport equation}

\author[]{Oleksandr Koshkarov}

\author[]{L. Chacón\corref{cor1}}
\ead{chacon@lanl.gov}

\cortext[cor1]{Corresponding author}

\address{Los Alamos National Laboratory, Los Alamos, NM 87545, USA}
\begin{abstract}
In this paper, we extend the operator-split asymptotic-preserving,
semi-Lagrangian algorithm for time dependent anisotropic heat transport
equation proposed in {[}Chacón et al., JCP, 272, 719-746, 2014{]}
to use a fully implicit time integration with backward differentiation
formulas. The proposed implicit method can deal with arbitrary heat-transport
anisotropy ratios $\chi_{\parallel}/\chi_{\perp}\ggg1$ (with $\chi_{\parallel}$,
$\chi_{\perp}$ the parallel and perpendicular heat diffusivities,
respectively) in complicated magnetic field topologies in an accurate
and efficient manner. The implicit algorithm is second-order accurate
temporally, has favorable positivity preservation properties, and
demonstrates an accurate treatment at boundary layers (e.g., island
separatrices), which was not ensured by the operator-split implementation.
The condition number of the resulting algebraic system is independent
of the anisotropy ratio, and is inverted with preconditioned GMRES.
We propose a simple preconditioner that renders the linear system
compact, resulting in mesh-independent convergence rates for topologically
simple magnetic fields, and convergence rates scaling as $\sim(N\Delta t)^{1/4}$
(with $N$ the total mesh size and $\Delta t$ the timestep) in topologically
complex magnetic-field configurations. We demonstrate the accuracy
and performance of the approach with test problems of varying complexity,
including an analytically tractable boundary-layer problem in a straight
magnetic field, and a topologically complex magnetic field featuring
magnetic islands with extreme anisotropy ratios $(\chi_{\parallel}/\chi_{\perp}=10^{10})$.
\end{abstract}
\maketitle

\section{Introduction}

\label{S:1} Understanding heat transport in magnetized plasmas is
important for many systems of interests such as magnetic fusion, space,
and astrophysical plasmas. Unfortunately, the magnetic field introduces
significant heat transport anisotropy, which prevents the use of standard
discretizations and solvers for parabolic/elliptic equations. In particular,
heat transport is significantly faster along the field line than across
it. For example, theoretical estimates, experimental measurements,
and modeling suggest that the transport anisotropy in common tokamak
reactors can reach extremely high values $\chi_{\parallel}/\chi_{\perp}\sim10^{7}$
--- $10^{10}$ \citep{braginskii1963transport,holzl2009determination,ren1998measuring,meskat2001analysis,snape2012influence,choi2014improved},
where $\chi_{\parallel}$ and $\chi_{\perp}$ are thermal conductivities
along and perpendicular to the magnetic field line, respectively.
Such high transport anisotropy presents significant difficulties for
its numerical integration. Common explicit methods would require one
to resolve the fastest time scale, which is prohibitive. At the same
time, implicit methods are challenged by the inversion of an almost
singular operator, with condition number $\kappa$ proportional to
the anisotropy ratio, $\kappa\sim\chi_{\parallel}/\chi_{\perp}$.
Moreover, tiny numerical errors in the discretization of the parallel
heat transport term, magnified by the anisotropy ratio, will seriously
pollute subtle perpendicular dynamics of dynamical importance. High-order
methods can be used to overcome error pollution \citep{sovinec2004nonlinear,gunter2005modelling,gunter2007finite,gunter2009mixed},
but they lack monotonicity and a maximum principle %\todo[fancyline]{is it enough for monotonicity comment?}
\citep{sharma2007preserving,kuzmin2009constrained}, and still lead
to nearly singular parallel transport operators. It is possible to
preserve monotonicity using limiters \citep{sharma2007preserving,dutoit2018positivity},
but at the expense of reverting to a low-order method.

Nonlocal heat closures, usually encountered in collisionless plasmas
\citep{held2001conductive,held2004nonlocal} and stochastic magnetic
fields in 3D, pose another challenge for conventional numerical methods.
Almost a perfect solution for all those problems (anisotropy, nonlocal
closures, and chaotic magnetic fields) in the limit $\nabla\cdot(\mathbf{B}/B)\approx0$
\footnote{In tokamak reactors, the quantity $\nabla\cdot(\mathbf{B}/B)=\nabla\cdot\mathbf{b}$
scales as inverse aspect ratio multiplied by the ratio of poloidal
to toroidal magnetic field, so the approximation $\nabla\cdot\mathbf{b}\approx0$
is particularly relevant for high-aspect-ratio, large-guide-field
tokamaks. Accordingly, it is often called \textquotedbl tokamak ordering\textquotedbl{}
approximation. %Moreover, if aspect ratio is not large, 
%this term may plays a minor role when the parallel transport is very fast,
%i.e.,
%$\nabla \cdot \mathbf b \ll \mathbf b \cdot \nabla$.
} was proposed for purely parallel transport \citep{del2011local,del2012parallel}
and then extended to include perpendicular transport \citep{chacon2014asymptotic}.
The key feature proposed in Refs. \citep{del2011local,del2012parallel,chacon2014asymptotic}
is a Green's function formulation of the heat transport equation,
where parallel fast transport is resolved essentially analytically.
Those methods ensure the absence of pollution due to their asymptotic
preserving nature \citep{chacon2014asymptotic,larsen1987asymptotic,larsen1989asymptotic,jin1999efficient}
in the limit of infinite anisotropy, and grant the ability to deal
with nonlocal heat closures and arbitrary magnetic field topology.
Unfortunately, the algorithm proposed in Ref. \citep{chacon2014asymptotic}
is based on a first-order operator splitting, and features an accuracy-based
time step limitation in the presence of boundary layers in the magnetic
field topology (e.g., island separatrices).

To remove this limitation, we propose in this study the extension
of this algorithm to allow an implicit time integration (the possibility
for such an extension was briefly formulated in \citep{chacon2014asymptotic}
in an appendix) with first- and second-order backward differentiation
formulas. We show by analysis that the condition number of the resulting
linear matrix is independent of the anisotropy ratio $\chi_{\parallel}/\chi_{\perp}$.
Preconditioned GMRES \citep{saad1986gmres} is used to invert the
associated algebraic system. GMRES is chosen (instead of CG) because,
while formally our system (for selected sets of boundary conditions)
is symmetric, we do not expect the semi-Lagrangian algorithm to remain
strictly symmetric due to the interpolations performed along fields
lines. Importantly, we propose a simple preconditioner that renders
the linear system compact, resulting in mesh-independent convergence
rates for topologically simple magnetic fields, and convergence rates
scaling as $\sim(N\Delta t)^{1/4}$ (with $N$ the total mesh size
and $\Delta t$ the timestep) in topologically complex magnetic-field
configurations. In many applications of interest (e.g., thermonuclear
magnetic fusion), the smallness of $\chi_{\perp}$ results in $\mathcal{O}(1)$
number of iterations, and therefore a practical algorithm.

Alternate AP schemes for the anisotropic transport equation have been
proposed in the literature, e.g., Refs. \citep{degond2010duality,degond2010asymptotic,degond2012asymptotic,narski2014asymptotic,wang2018uniformly}.
Refs. \citep{degond2010duality,degond2010asymptotic,degond2012asymptotic}
considered only open field lines in a time-independent context. In
contrast, Ref. \citep{wang2018uniformly} considered only closed ones,
also in a time-independent context. Ref. \citep{narski2014asymptotic}
considered the time-dependent case for open and closed magnetic fields
with implicit timestepping. However, it is unclear how the approach
can generalize to three dimensions (where confined stochastic field
lines of infinite length may exist), and the reference employed a
direct linear solver, which is known to scale very poorly with mesh
refinement and with processor count in parallel environments. In contrast,
our approach can be readily extended to 3D (see Refs. \citep{del2011local,del2012parallel}
for 3D computations in the purely parallel transport case), is easily
parallelizable, and scales much better with mesh refinement than either
direct or unpreconditioned Krylov iterative methods.

The rest of the paper is organized as follows. In Section~\ref{sec:formulation},
we summarize the Lagrangian Green's function formalism, the cornerstone
of the proposed method. Next, we describe the implicit temporal discretization
in Section~\ref{sec:time}. In Section~\ref{sec:spec}, the spectral
properties of the method are analyzed, followed by the analysis of
positivity in Section~\ref{sec:positivity}. In Section~\ref{sec:prec},
we formulate our preconditioning strategy. Numerical implementation
details are provided in Section~\ref{sec:impl}. Numerical tests
demonstrating the merits of the new method are provided in Section~\ref{sec:ntests}.
Finally, discussion and conclusions are provided in Sections~\ref{sec:dis}
and \ref{sec:con}, respectively.

%%%%%%%%%%%%%%%%%%%%%%%%%%%%%%%%%%%%%%%%%%%%%%%%%%%%%%%
%%%%%  Problem formulation                        %%%%%
%%%%%%%%%%%%%%%%%%%%%%%%%%%%%%%%%%%%%%%%%%%%%%%%%%%%%%%

\section{The Lagrangian Green's function formulation}

\label{sec:formulation} For simplicity, we consider the simplest
anisotropic temperature transport equation, later revising the assumptions
that can be straightforwardly dropped. The anisotropic transport equation,
normalized to the perpendicular transport time and length scales ($L_{\perp}$,
$\tau_{\perp}=L_{\perp}^{2}/\chi_{\perp}$), reads: 
\begin{equation}
\partial_{t}T-\frac{1}{\epsilon}\nabla_{\parallel}^{2}T=\nabla_{\perp}^{2}T+S\equiv S_{*},\label{diffusion_eq}
\end{equation}
where $T=T(t,\mathbf{x})$ is a temperature profile, $S=S(t,\mathbf{x})$
is a heat source, $S_{*}=\nabla_{\perp}^{2}T+S$ is a formal source
to the purely parallel transport equation, $\epsilon=\tau_{\parallel}/\tau_{\perp}=(L_{\parallel}^{2}/\chi_{\parallel})/(L_{\perp}^{2}/\chi_{\perp})$
is the ratio between parallel and perpendicular (to the magnetic field)
transport time scales, with $L,\chi$ being spatial normalization
length scale and thermal conductivity, respectively. The thermal conductivity
along and perpendicular to the magnetic field ($\chi_{\parallel}$
and $\chi_{\perp}$) are assumed to be constants. The generalization
to non-constant conductivities is left for future work. The equation
needs to be supplemented with problem-dependent boundary and initial
conditions. Temperature boundary conditions at non-periodic boundaries
can in principle be arbitrary. Unless otherwise stated, we consider
homogeneous Dirichlet boundary conditions.

The differential operators along and perpendicular to the magnetic
field $\mathbf{B}=\mathbf{b}B$ are defined as 
\[
\nabla_{\parallel}^{2}=\nabla\cdot(\mathbf{b}\mathbf{b}\cdot\nabla)=\left(\cancelto{\approx0}{(\nabla\cdot\mathbf{b})}+(\mathbf{b}\cdot\nabla)\right)(\mathbf{b}\cdot\nabla)\approx(\mathbf{b}\cdot\nabla)^{2},\quad\nabla_{\perp}^{2}=\nabla^{2}-\nabla_{\parallel}^{2},
\]
where $\nabla\cdot\mathbf{b}\approx0$ is assumed. The topology of
magnetic fields can otherwise be arbitrary (including stochastic magnetic
fields). However, for the purpose of this paper, we will assume periodic
boundary conditions or perfectly confined magnetic fields, i.e., $\mathbf{n}\cdot\mathbf{b}=0$
at boundaries, where $\mathbf{n}$ is the unit vector normal to the
domain boundary. 

Next, we consider the analytical solution of the anisotropic transport
equation with the Green's function formalism \citep{chacon2014asymptotic}.
The formal implicit solution with initial condition $T(0,\mathbf{x})=T_{0}(\mathbf{x})=T_{0}$
is: 
\begin{align}
T(t,\mathbf{x})=\mathcal{G}\left(T_{0};\mathbf{x},\frac{t}{\epsilon}\right)+\int_{0}^{t}dt'\mathcal{G}\left(S_{*};\mathbf{x},\frac{t-t'}{\epsilon}\right),\label{green:sol}
\end{align}
where 
\begin{equation}
\mathcal{G}\left(T_{0};\mathbf{x},t\right)=\int dsG(s,t)T_{0}\left(\mathbf{\hat{x}}(s,\mathbf{x})\right)\label{eq:prop}
\end{equation}
is the propagator of the homogeneous transport equation, and $G(s,t)$
is the Green's function of the diffusion equation (see below). The
integration in equation (\ref{eq:prop}) is performed along the magnetic
field line, which passes through $\mathbf{x}$ and is parameterized
by the arc length~$s$, 
\begin{align}
\frac{d\mathbf{\hat{x}}(s)}{ds}=\mathbf{b},\quad\mathbf{\hat{x}}(0)=\mathbf{x}.\label{trace_mag}
\end{align}
In the case of perfectly confined magnetic field lines (which start
and end at infinity), the Green's function takes the form 
\begin{equation}
G(s,t)=\frac{1}{\sqrt{4\pi t}}\exp\left(-\frac{s^{2}}{4t}\right).\label{green:fun}
\end{equation}
In principle, finite magnetic-field lines and/or nonlocal heat closures
can be easily incorporated by providing the appropriate Green's function
without modifying the analysis \citep{chacon2014asymptotic}.

As was shown in previous studies \citep{del2011local,del2012parallel,chacon2014asymptotic},
the Lagrangian formulation (\ref{green:sol}) features key properties,
namely, that $\mathcal{G}$ is the identity on the null space of the
parallel diffusion operator $\nabla_{\parallel}^{2}$ and that $\lim_{t\to\infty}\mathcal{G}$
is the projector onto that null space. These play a central role in
controlling numerical pollution, and ensuring the asymptotic preserving
properties of any numerical method constructed based on equation (\ref{green:sol})
when $\epsilon\rightarrow0$. We construct an implicit time discretization
of equation (\ref{green:sol}) in the next section.

%%%%%%%%%%%%%%%%%%%%%%%%%%%%%%%%%%%%%%%%%%%%%%%%%%%%%%%
%%%%%  Implicit time discretization               %%%%%
%%%%%%%%%%%%%%%%%%%%%%%%%%%%%%%%%%%%%%%%%%%%%%%%%%%%%%%

\section{Implicit time discretization and asymptotic-preserving property}

\label{sec:time} We follow Ref.~\citep{chacon2014asymptotic} and
begin by constructing an implicit Euler time discretization, the first-order
backward differentiation formula (BDF1). First, we rewrite equation
(\ref{green:sol}) for a chosen time step with $t^{n}=n\Delta t$
and $T^{n}(\mathbf{x})=T(t^{n},\mathbf{x})$\textcolor{red}{{} }\textcolor{black}{
\begin{align}
T^{n+1}(\mathbf{x})=\mathcal{G}\left(T^{n};\mathbf{x},\frac{\Delta t}{\epsilon}\right)+\int_{t^{n}}^{t^{n+1}}dt'\mathcal{G}\left(S_{*};\mathbf{x},\frac{t^{n+1}-t'}{\epsilon}\right).
\end{align}
}Next, in order to evaluate the time integral, we assume the implicit
source term is constant during the time step, $S_{*}(t)\approx S_{*}(t^{n+1})\equiv S_{*}^{n+1}$,
so 
\begin{align}
\int_{t^{n}}^{t^{n+1}}dt'\mathcal{G}\left(S_{*};\mathbf{x},\frac{t^{n+1}-t'}{\epsilon}\right)=\Delta t\mathcal{P}\left(S_{*}^{n+1};\mathbf{x},\frac{\Delta t}{\epsilon}\right)+\mathcal{E}_{src}^{BDF1},\label{time_int}
\end{align}
where 
\begin{align}
\mathcal{P}\left(S_{*}^{n+1};\mathbf{x},\frac{\Delta t}{\epsilon}\right)=\int_{-\infty}^{+\infty}ds\mathcal{U}\left(s,\frac{\Delta t}{\epsilon}\right)S_{*}^{n+1}\left(\mathbf{\hat{x}}(s,\mathbf{x})\right),\label{prop_def}
\end{align}
and 
\begin{align}
\mathcal{U}\left(s,\frac{\Delta t}{\epsilon}\right)=\frac{1}{\Delta t}\int_{t^{n}}^{t^{n+1}}dt'G\left(s,\frac{t^{n+1}-t'}{\epsilon}\right).
\end{align}
For the Green function (\ref{green:fun}), $\mathcal{U}$ takes the
form: 
\[
\mathcal{U}\left(s,\tau\right)=\frac{1}{\sqrt{\tau}}\left(\frac{e^{-s^{2}/4\tau}}{\sqrt{\pi}}-\frac{|s|}{2\sqrt{\tau}}\erfc\left(\frac{|s|}{2\sqrt{\tau}}\right)\right),
\]
where $\erfc(x)$ is the complementary error function. The equation
(\ref{time_int}) introduces the BDF1 local discretization error,
which as shown in Ref. \citep{chacon2014asymptotic} is second-order
$\mathcal{E}_{src}^{BDF1}=O(\Delta t^{2})$, confirming that BDF1
is globally a first-order method. As a result, we obtain the implicit
BDF1 discretization sought: 
\begin{align}
T(\mathbf{x})^{n+1}=\mathcal{G}\left(T^{n};\mathbf{x},\frac{\Delta t}{\epsilon}\right)+\Delta t\mathcal{P}\left(S_{*}^{n+1};\mathbf{x},\frac{\Delta t}{\epsilon}\right)+\mathcal{E}_{src}^{BDF1},\label{bdf1}
\end{align}
which can be used further to construct higher-order BDF formulas.
For the purpose of this paper and to illustrate the general procedure,
we derive the second-order BDF, i.e., BDF2. Rewriting equation (\ref{bdf1})
for the time step~$2\Delta t$:
\begin{align}
T(\mathbf{x})^{n+1}=\mathcal{G}\left(T^{n-1};\mathbf{x},\frac{2\Delta t}{\epsilon}\right)+2\Delta t\mathcal{P}\left(S_{*}^{n+1};\mathbf{x},\frac{2\Delta t}{\epsilon}\right)+4\mathcal{E}_{src}^{BDF1},\label{bdf1x2}
\end{align}
and subtracting equations (\ref{bdf1}), (\ref{bdf1x2}) such that
the leading-order error terms cancel, we obtain:
\begin{align}
\begin{split}T(\mathbf{x})^{n+1} & =\frac{4}{3}\mathcal{G}\left(T^{n};\mathbf{x},\frac{\Delta t}{\epsilon}\right)-\frac{1}{3}\mathcal{G}\left(T^{n-1};\mathbf{x},\frac{2\Delta t}{\epsilon}\right)\\
 & +\frac{2\Delta t}{3}\left[2\mathcal{P}\left(S_{*}^{n+1};\mathbf{x},\frac{\Delta t}{\epsilon}\right)-\mathcal{P}\left(S_{*}^{n+1};\mathbf{x},\frac{2\Delta t}{\epsilon}\right)\right]+\mathcal{E}_{src}^{BDF2}.
\end{split}
\label{bdf2}
\end{align}
The expressions for truncation errors $\mathcal{E}_{src}^{BDF1}$
and $\mathcal{E}_{src}^{BDF2}$ were derived exactly for straight
magnetic field lines in Ref.~\citep{chacon2014asymptotic}, and their
Fourier amplitudes are:
\begin{align}
 & \mathcal{\hat{E}}_{src}^{BDF1}(k_{\parallel},k_{\perp})=\hat{T}(k_{\parallel},k_{\perp})\min\left[\left(\frac{\Delta t}{\tau_{k}}\right)^{2},\frac{\epsilon}{k_{\parallel}^{2}\Delta t}\right],\label{err:bdf1}\\
 & \mathcal{\hat{E}}_{src}^{BDF2}(k_{\parallel},k_{\perp})=\hat{T}(k_{\parallel},k_{\perp})\min\left[\left(\frac{\Delta t}{\tau_{k}}\right)^{2}O(\Delta t),\frac{\epsilon}{k_{\parallel}^{2}\Delta t}\right],\label{err:bdf2}
\end{align}
with 
\begin{align}
\frac{\Delta t}{\tau_{k}}=\min\left[\Delta t\left(k_{\perp}^{2}+k_{\parallel}^{2}/\epsilon+\frac{\hat{S}(k_{\parallel},k_{\perp})}{\hat{T}(k_{\parallel},k_{\perp})}\right),1\right],
\end{align}
where $\hat{T}(k_{\parallel},k_{\perp})$ and $\hat{S}(k_{\parallel},k_{\perp})$
are Fourier components of the temperature field and the external heat
source, respectively. Notice that errors in equations (\ref{err:bdf1}),
(\ref{err:bdf2}) are either proportional or independent of $\epsilon$,
which guarantees a bounded error for high anisotropy, $\epsilon\ll1$,
and implying that the formulation is asymptotic preserving.

\section{Spectral analysis}

\label{sec:spec} In this section, we analyze the spectral behavior
of the resulting scheme, which directly depends on the spectral properties
of $\mathcal{G}$ and $\mathcal{P}$. We begin by deriving the Fourier
transform of the propagation operators. Following \citep{chacon2014asymptotic},
we find: 
\begin{align}
\int ds\mathcal{G}(f;\mathbf{x},t)e^{-ik_{\parallel}s}=\hat{f}_{k_{\parallel}}e^{-k_{\parallel}^{2}t},\label{fop:g}
\end{align}
and 
\begin{align}
\int ds\mathcal{P}(f;\mathbf{x},t)e^{-ik_{\parallel}s}=\hat{f}_{k_{\parallel}}\frac{1-e^{-k_{\parallel}^{2}t}}{k_{\parallel}^{2}t}.\label{fop:p}
\end{align}

\subsection{Spectral properties of the linear operator $\mathcal{L}$}

For the sake of brevity, we simplify notations in this section and
rewrite equation (\ref{bdf1}) in the form 
\begin{align}
\mathcal{L}T=v,\quad\mbox{ with }\quad\mathcal{L}=\mathbb{I}+\mathcal{P}\mathbb{B},\label{lin_system}
\end{align}
where $T=T^{n+1}(\mathbf{x})$, $v$ is the right hand side that gathers
contributions from terms independent of $T$ (previous time step and
heat source terms), $\mathcal{P}f=\mathcal{P}(f;\mathbf{x},\Delta t/\epsilon)$,
and 
\begin{align}
\mathbb{B}=-\Delta t\nabla_{\perp}^{2}.\label{eq:beta-def}
\end{align}
We analyze the spectrum of the operator $\mathcal{L}$ by Fourier-analyzing
(which is only rigorous in the straight magnetic field case with $\mathbf{b}$
aligned with a mesh coordinate), and using expressions (\ref{fop:g})
and (\ref{fop:p}), giving the eigenvalues:
\[
\lambda=1+\frac{\beta}{\alpha}\left(1-e^{-\alpha}\right),
\]
where we have introduced ``conductivity scales'' for parallel and
perpendicular directions as:
\begin{equation}
\alpha=\frac{1}{\epsilon}\Delta tk_{\parallel}^{2},\quad\beta=\Delta tk_{\perp}^{2},
\end{equation}
respectively. Note that $\beta$ is the eigenvalue of $\mathbb{B}$
operator, Eq. (\ref{eq:beta-def}). It follows that, for components
in the null space of the parallel transport operator ($k_{\parallel}=0$),
we have:
\[
\lambda_{k_{\parallel}=0}=1+\Delta tk_{\perp}^{2}.
\]
Otherwise, we get:
\[
\lambda_{k_{\parallel}\neq0}=1+\mathcal{O}(\epsilon),
\]
which vanishes in the limit of arbitrarily small $\epsilon$. It follows
the condition number of the system matrix $\mathcal{L}$ is: 
\begin{equation}
\kappa_{\mathcal{L}}=\frac{\lambda_{max}}{\lambda_{min}}=1+\Delta tk_{\perp}^{2}+\mathcal{O}(\epsilon),\label{eq:L_cond_number}
\end{equation}
which is largely independent of the small parameter $\epsilon$, and
so will be the performance of the iterative method inverting it. This
will be verified numerically in Sec. \ref{sec:ntests}.

\subsection{Linear stability}

We establish next the stability of the BDF1 and BDF2 discretizations
in the Von-Neumann sense. One can perform a Von Neumann temporal stability
analysis, by using the usual ansatz 
\begin{equation}
T^{n}\sim\sigma^{n}(k_{\parallel},k_{\perp})e^{i\mathbf{x}\cdot\mathbf{k}}.\label{vn}
\end{equation}
The straightforward application of (\ref{vn}) to (\ref{bdf1}) and
(\ref{bdf2}), using expressions (\ref{fop:g}) and (\ref{fop:p}),
leads to 
\begin{equation}
\sigma=e^{-\alpha}-\sigma\frac{\beta}{\alpha}\left(1-e^{-\alpha}\right),
\end{equation}
for BDF1, and 
\begin{equation}
\sigma=\frac{4}{3}e^{-\alpha}-\frac{1}{3\sigma}e^{-2\alpha}-\sigma\frac{\beta}{\alpha}\left(1-\frac{4}{3}e^{-\alpha}+\frac{1}{3}e^{-2\alpha}\right),
\end{equation}
for BDF2. Absolute stability requires $|\sigma|<1$, which in turn
demands that 
\begin{equation}
\left|1+\frac{\beta}{\alpha}\left(1-e^{-\alpha}\right)\right|>e^{-\alpha},\label{bdf1_st}
\end{equation}
for BDF1, and 
\begin{align}
\left|2\pm\sqrt{1-3\frac{\beta}{\alpha}\left(1-\frac{4}{3}e^{-\alpha}+\frac{1}{3}e^{-2\alpha}\right)}\right|>e^{-\alpha},\label{bdf2_st}
\end{align}
for BDF2. We can see that equations (\ref{bdf1_st}) and (\ref{bdf2_st})
are always true, as the left hand side is always greater than unity,\footnote{Because $\left|2\pm\sqrt{1-a}\right|>1$, for $a>0$}
while the right hand side is always less then unity. Hence, BDF1 and
BDF2 are absolutely stable for arbitrary $\alpha$ and $\beta$ (i.e.,
arbitrary $\Delta t$, $k_{\parallel}$, $k_{\perp}$, and $\epsilon$).

%%%%%%%%%%%%%%%%%%%%%%%%%%%%%%%%%%%%%%%%%%%%%%%
%%%%%  Convergence of Krylov method       %%%%%
%%%%%%%%%%%%%%%%%%%%%%%%%%%%%%%%%%%%%%%%%%%%%%%

\section{Positivity}

\label{sec:positivity} The continuum heat transport equation preserves
positivity of the temperature field in the absence of sources and
sinks. In this section, we show that our implicit temporal discretization
applied to semi-Lagrangian formulation has robust positivity preservation
properties given realistic temperature initial conditions. \textcolor{black}{In
order to carry out the analysis, we consider the semi-discrete context
(discrete in time and continuum in space), and ignore important details
of the Eulerian discretization of the isotropic Laplacian operator.
In practice, positivity preservation by the semi-Lagrangian scheme
will require (at the very least) having a discrete representation
of the Laplacian that features a maximum principle.} We assume a straight
magnetic field and an infinite domain (to allow a Fourier transform).
We expect that the disregarded magnetic-field curvature effects will
not significantly change the positivity properties of the scheme,
owing to the temperature flattening effect of fast parallel transport.
Additionally, the current analysis assumes that all field lines are
contained in the computational domain, thus avoiding dealing with
problem-dependent boundary conditions. For the purpose of this section,
we consider only the BDF1 method.

Starting with the Fourier transform of (\ref{bdf1}), we get: 
\begin{equation}
\hat{T}^{n+1}=\frac{e^{-\alpha}}{1+\beta(\frac{1-e^{-\alpha}}{\alpha})}\hat{T}^{n},
\end{equation}
where $\hat{T}$ is a Fourier transform of the temperature field.
Using the convolution theorem, we can write: 
\begin{equation}
T^{n+1}(\mathbf{x})=\int d^{3}\mathbf{x}'\mathcal{K}(\mathbf{x}-\mathbf{x}')T^{n}(\mathbf{x}'),\label{exact_step_kernel}
\end{equation}
where the kernel is 
\begin{equation}
\mathcal{K}(\mathbf{x})=\frac{1}{(2\pi)^{3}}\text{Re}\left[\int d^{3}ke^{i\mathbf{k}\cdot\mathbf{x}}\frac{e^{-\alpha}}{1+\beta(\frac{1-e^{-\alpha}}{\alpha})}\right],
\end{equation}
and the real part is taken for convenience (because the imaginary
part is trivially zero by symmetry). After a straightforward manipulation
(see \ref{sec:kernel}), one can simplify the integral to 
\begin{align}
\mathcal{K}(\mathbf{x})=\frac{1}{4\pi^{2}\Delta t}\int_{-\infty}^{\infty}dk_{\parallel}\cos(k_{\parallel}z)e^{-\alpha}\frac{\alpha}{1-e^{-\alpha}}K_{0}\left(r_{\perp}\sqrt{\frac{\alpha}{1-e^{-\alpha}}}\right),
\end{align}
where $r_{\perp}=|\mathbf{x}-\mathbf{b}\mathbf{b}\cdot\mathbf{x}|$,
$z=\mathbf{b}\cdot\mathbf{x}$ is the coordinate along the magnetic
field, and $K_{0}$ is the modified Bessel function of the second
kind. Notice that $K_{0}(0)=\infty$, and therefore the kernel is
singular when $r_{\perp}\rightarrow0$. However, this singularity
is integrable, since $K_{0}(x)\sim-\ln(x)$ for $x\sim+0$. Physically,
the singularity at $r_{\perp}\rightarrow0$ means that most of the
contribution to $T^{n+1}(\mathbf{x}_{0})=T^{n+1}(x_{0},y_{0},z_{0})$
(where $x$ and $y$ are coordinates perpendicular to the magnetic
field) will come from the previous time step temperature along the
magnetic field line passing through $(x_{0},y_{0},z_{0})$, $T^{n}(x_{0},y_{0},z')$.
Also note that $\mathcal{K}(\mathbf{x})$ decays quickly for large
$r_{\perp}$, since $K_{0}(x)\sim e^{-x}/\sqrt{x}$ for $x\rightarrow\infty$.

\begin{figure}[htb]
\centering \begin{subfigure}[t]{60mm} \includegraphics[width=1\linewidth]{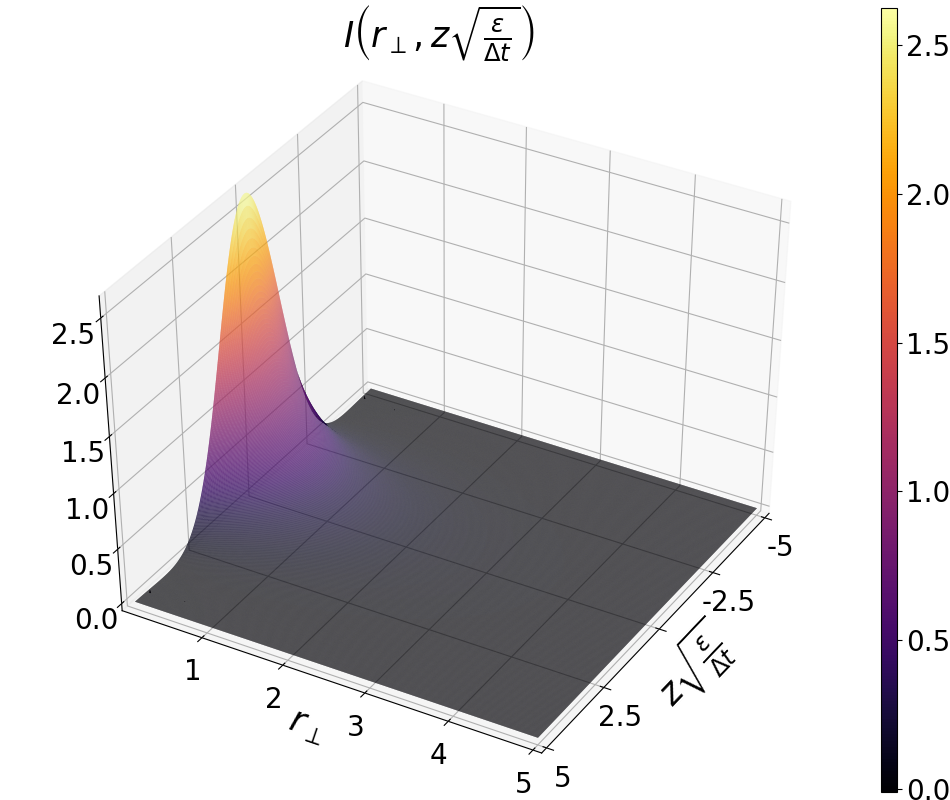}
\caption{$0.1<r_{\perp}<5$, $-5<z\sqrt{\epsilon/\Delta t}<5$.}
\label{fig:pos:I:full} \end{subfigure} \begin{subfigure}[t]{60mm}
\includegraphics[width=1\linewidth]{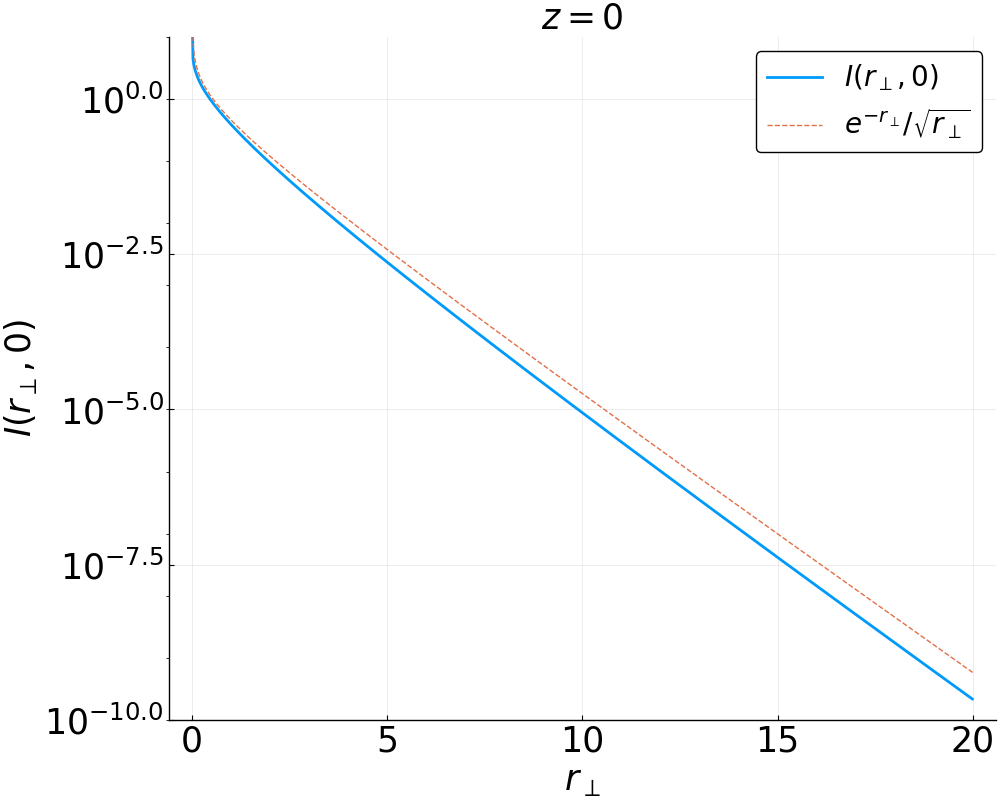} \caption{$10^{-4}<r_{\perp}<20$, $z=0$.}
\label{fig:pos:I:r} \end{subfigure} \begin{subfigure}[t]{60mm}
\includegraphics[width=1\linewidth]{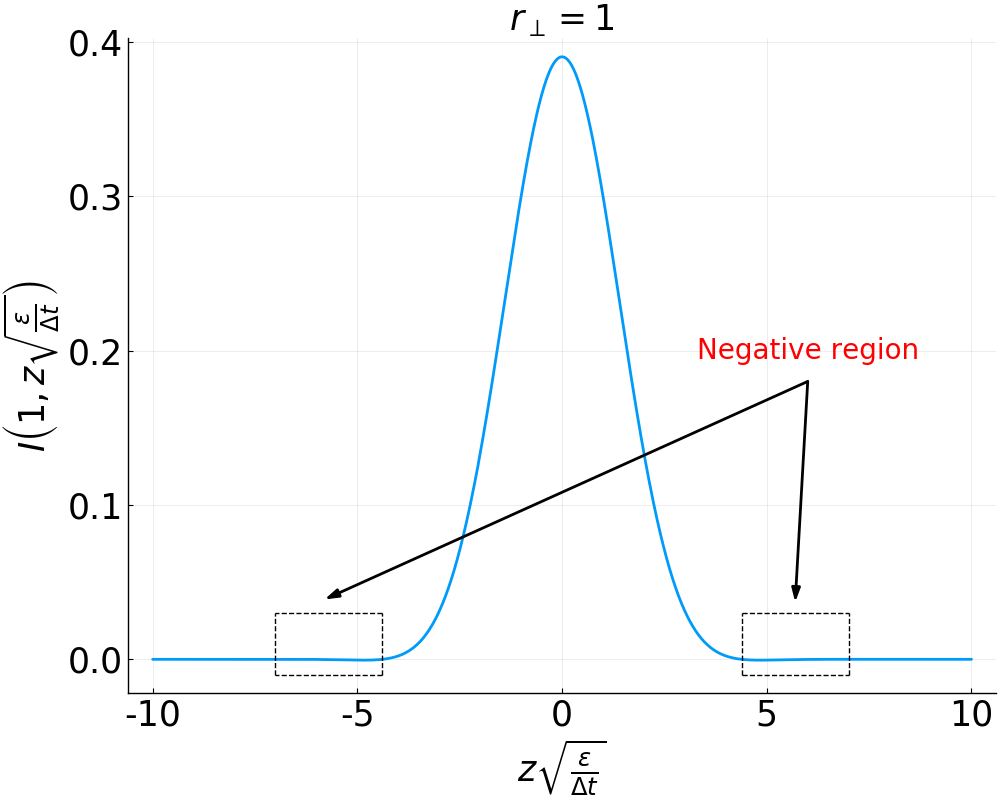} \caption{$r_{\perp}=1$, $-10<z\sqrt{\epsilon/\Delta t}<10$.}
\label{fig:pos:I:z} \end{subfigure} \begin{subfigure}[t]{60mm}
\includegraphics[width=1\linewidth]{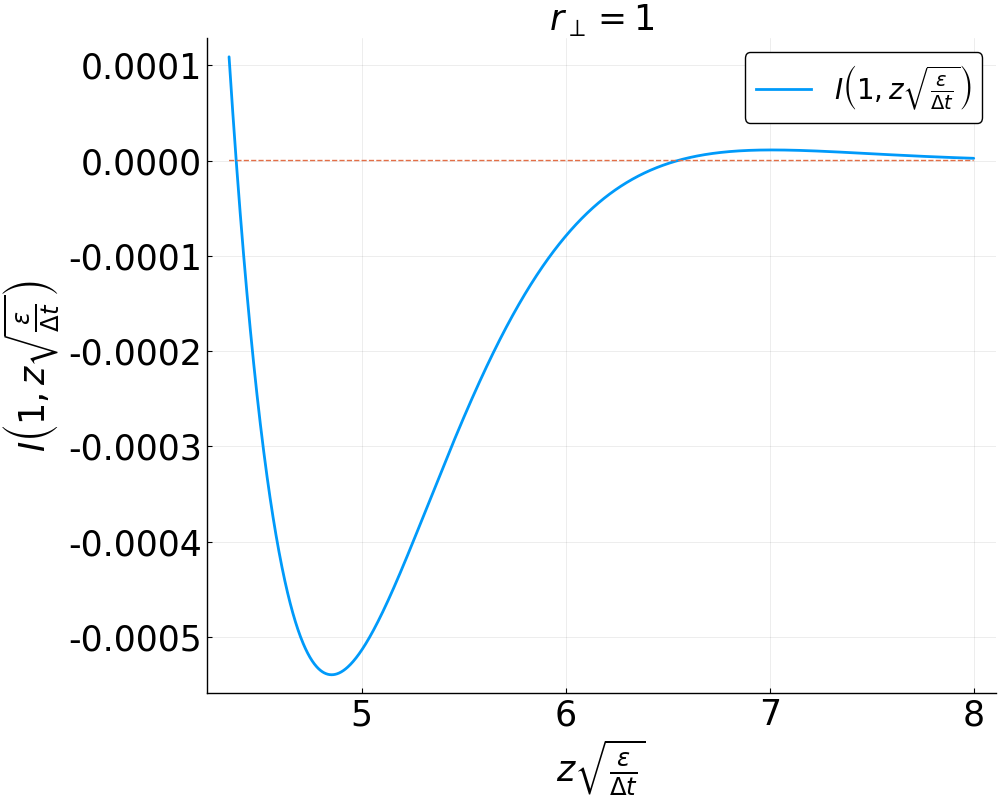} \caption{$r_{\perp}=1$, $4.35<z\sqrt{\epsilon/\Delta t}<8$.}
\label{fig:pos:I:z:zoom} \end{subfigure} \caption{Numerically evaluated integral kernel (\ref{pos:I})}
\label{fig:pos:I}
\end{figure}

To study positivity, we write the kernel $\mathcal{K}$ in the form
\begin{equation}
\mathcal{K}(\mathbf{x})=\frac{\sqrt{\epsilon}}{2\pi^{2}(\Delta t)^{3/2}}I\left(r_{\perp},z\sqrt{\frac{\epsilon}{\Delta t}}\right),
\end{equation}
with 
\begin{equation}
I(a,b)=\int_{0}^{\infty}d\xi\cos(\xi b)e^{-\xi^{2}}\frac{\xi^{2}}{1-e^{-\xi^{2}}}K_{0}\left(a\frac{\xi}{\sqrt{1-e^{-\xi^{2}}}}\right),\label{pos:I}
\end{equation}
so the positivity of $T^{n+1}$ depends only on the integral $I$,
which depends on two parameters: $r_{\perp}$ and $z\sqrt{\epsilon/\Delta t}$.
The numerically obtained $I$ is shown in Figure~\ref{fig:pos:I:full}.
The main feature of $I$ is an infinite logarithmic peak at $r_{\perp}=z=0$
(or $x=y=z=0$) and fast decay in all other directions. Perpendicular
to the magnetic field, the kernel damps monotonically $\mathcal{K}(r_{\perp})\sim I(r_{\perp})\sim K_{0}(r_{\perp})\sim e^{-r_{\perp}}/\sqrt{r_{\perp}}$
as expected and as shown in Figure~\ref{fig:pos:I:r}. The kernel
also quickly damps in the direction along magnetic field (when $z$
increases), as shown in Figure~\ref{fig:pos:I:z}.

If the kernel were positive everywhere, then $T^{n+1}$ would be guaranteed
to be positive for any given positive $T^{n}$. However, the kernel
$\mathcal{K}$ can be slightly negative in a small region around $z\sim5\sqrt{\Delta t/\epsilon}$,
as shown in Figure~\ref{fig:pos:I:z:zoom} (zoom of Figure~\ref{fig:pos:I:z}).
Therefore, $T^{n+1}$ can in principle be negative for some unphysical
temperature profile when, for example, temperature is zero everywhere
except around $z=5\sqrt{\Delta t/\epsilon}$ (which can be a very
distant region along the magnetic field line for $\epsilon\ll1$).
However, in most practical situations, the temperature field will
have the form \citep{chacon2014asymptotic}: 
\begin{equation}
T^{n}=\langle T^{n}\rangle+T_{\epsilon}^{n},
\end{equation}
where $\langle T^{n}\rangle$ is a constant component along magnetic
field lines (due to fast conductivity along magnetic field lines),
and $T_{\epsilon}^{n}\sim\epsilon$ is a small correction. Therefore,
in practice, the convolution with kernel $\mathcal{K}$ will always
be positive, which ensures the positivity of the solution.

\section{Preconditioning strategy}

\label{sec:prec} Equations (\ref{bdf1}) and (\ref{bdf2}) are implicit.
T\textcolor{black}{hus, one needs to invert them at every time step.
However, a matrix representation for any given discretization of the
operators $\mathcal{G}$ and $\mathcal{P}$ for arbitrary $\mathbf{B}$
is not known beforehand, and would be extremely difficult to construct.
Therefore, it is advantageous to use robust matrix-free Krylov methods
such as GMRES (generalized minimal residuals) \citep{saad1986gmres}.
}

\textcolor{black}{In general, the convergence of Krylov methods is
sensitive to the condition number of the linear operator. Indeed,
the number of GMRES (or CG for SPD systems) iterations scales as $N_{i}\sim\sqrt{\kappa}$
\citep{daniel1967conjugate}. However, for compact operators stemming
from integral equations, this performance estimate can be quite pessimistic,
and in practice performance of GMRES in particular can be found to
be mesh independent \citep{campbell1996convergence}. Our preconditioning
strategy is to render the right-preconditioned linear operator $\mathcal{L}\mathcal{M}^{-1}=(\mathbb{I}+\mathcal{P}\mathbb{B})\mathcal{M}^{-1}$
compact, such that all eigenvalues are included in the unit sphere,
and such that cluster around zero and one for sufficiently large timesteps.
Under such conditions, one may expect much faster Krylov convergence
than $\sqrt{\kappa}$ without further preconditioning \citep{campbell1996gmres}.}

\textcolor{black}{We propose the simple preconditioner (which tackles
perpendicular or isotropic transport only):
\[
\mathcal{M}^{-1}=(\mathbb{I}+\mathbb{B})^{-1}.
\]
When $\mathcal{M}^{-1}$ is applied as a right preconditioner (which
avoids subtleties with the non-commutativity between $\mathcal{P}$
and $\mathbb{B}$ in general magnetic-field configurations), we readily
find:
\begin{equation}
\mathcal{L}\mathcal{M}^{-1}=(\mathbb{I}+\mathcal{P}\mathbb{B})(\mathbb{I}+\mathbb{B})^{-1}=(\mathbb{I}+\mathbb{B})^{-1}+\mathcal{P}\mathbb{B}(\mathbb{I}+\mathbb{B})^{-1}.\label{eq:pc_op}
\end{equation}
In straight magnetic field topologies aligned with the mesh, one can
perform a Fourier analysis of the spectrum of the preconditioned operator.
The eigenvalues of $\mathcal{L}\mathcal{M}^{-1}$ can be found as:
\begin{equation}
\rho(\mathcal{L}\mathcal{M}^{-1})\approx\frac{1}{1+\beta}+\frac{1-e^{-\alpha}}{\alpha}\frac{\beta}{1+\beta}\leq1,\label{eq:compactness}
\end{equation}
for arbitrary $\alpha,\beta\in\mathbb{R}_{0}^{+}$. In Eq. \ref{eq:compactness},
equality is found for the null space ($\alpha=0$), while for off-null-space
components ($\alpha\gg1$) we find: 
\[
\rho_{\mathrm{min}}(\mathcal{L}\mathcal{M}^{-1})\overset{\alpha\gg1}{\longrightarrow}\frac{1}{1+\beta}\overset{\beta\gg1}{\longrightarrow}0.
\]
Thus the preconditioned operator $\mathcal{L}\mathcal{M}^{-1}$ is
compact, with $\rho\leq1$ and two clusters: one at $\rho=1^{-}$
(for null space components) and another at $\rho=0^{+}$ (for sufficiently
large $\beta$). This result is rigorous for straight magnetic fields
aligned with a mesh coordinate, as the operators $\mathcal{P}$ and
$\mathbb{B}$ commute.}

\textcolor{black}{In practice, straight magnetic fields are not of
much practical value. In topologically simple magnetic fields (e.g.,
without magnetic islands), we expect the temperature field to reside
in the null space of the parallel transport operator (corresponding
to the $\rho=1^{-}$ eigenvalue cluster), and convergence may be expected
to be mesh independent due to strong clustering of the corresponding
eigenvalues (see Chap. 7 of Ref. \citep{rasmussen2001compact} for
an analysis of the impact of the right hand side spectral content
on GMRES convergence). However, for topologically complex magnetic
fields, coupling of $\mathcal{P}$ and $\mathbb{B}$ operators at
boundary layers (e.g., island separatrices) will result in eigenvalue
spread and temperature fields with components orthogonal to the null
space.}

\textcolor{black}{For such complex magnetic field topologies, the
preconditioned operator is formally still ill-conditioned, with condition
number:
\[
\kappa=\frac{\rho_{max}}{\rho_{min}}\approx\beta_{\mathrm{max}}\sim\frac{\Delta t}{\Delta^{2}}\sim\Delta tN.
\]
Here, $\Delta$ is the characteristic cell size of the computational
grid, and $N\sim\Delta^{-2}$ is the total number of mesh points in
two dimensions. The expected number of Krylov iterations is expected
to scale as: 
\[
N_{it}\sim\sqrt{\kappa}\sim\sqrt{\beta_{\max}}\sim\sqrt{N\Delta t},
\]
which, while independent of $\alpha$ (and therefore $\epsilon$),
remains mesh-dependent and therefore formally not scalable. However,
the strong clustering of null space components almost surely implies
this prediction is too pessimistic. In fact, for large enough timesteps
such that $\beta_{\mathrm{min}}>1$, all the eigenvalues will cluster
either right below unity (for null-space components) or above zero
(for off-null-space ones) regardless of magnetic topology (because
$\mathbb{B}(\mathbb{I}+\mathbb{B})^{-1}\approx\mathbb{I}$ and therefore
$\mathcal{P}$ and $\mathbb{B}$ decouple in Eq. \ref{eq:pc_op}),
resulting in mesh-independent convergence rates. This will be borne
out by our numerical experiments. For $\beta_{min}<1$, the eigenvalue
clustering will degrade and the convergence rate will pick up a mesh
dependence, but we will find it numerically to scale as $(N\Delta t)^{1/4}$
instead of $(N\Delta t)^{1/2}$, which is significantly more favorable.}

\textcolor{black}{}

%%%%%%%%%%%%%%%%%%%%%%%%%%%%%%%%%%%%%%%%%%%%%%%%%%%%%%%
%%%%%  short note on implementation               %%%%%
%%%%%%%%%%%%%%%%%%%%%%%%%%%%%%%%%%%%%%%%%%%%%%%%%%%%%%%

\section{Numerical implementation details}

\label{sec:impl}

\subsection{Solver implementation details}

Given $T^{n}$, $T^{n-1}$, we seek the solution $T^{n+1}$ to the
BDF1 linear system:
\begin{align}
T(\mathbf{x})^{n+1}=\mathcal{G}\left(T^{n};\mathbf{x},\frac{\Delta t}{\epsilon}\right)+\Delta t\mathcal{P}\left(\nabla_{\perp}^{2}T^{n+1}+S^{n+1};\mathbf{x},\frac{\Delta t}{\epsilon}\right),\label{eq:bdf1-linear-system}
\end{align}
or, for BDF2:
\begin{align}
\begin{split}T^{n+1}(\mathbf{x}) & =\frac{4}{3}\mathcal{G}\left(T^{n};\mathbf{x},\frac{\Delta t}{\epsilon}\right)-\frac{1}{3}\mathcal{G}\left(T^{n-1};\mathbf{x},\frac{2\Delta t}{\epsilon}\right)\\
 & +\frac{2\Delta t}{3}\left[2\mathcal{P}\left(\nabla_{\perp}^{2}T^{n+1}+S^{n+1};\mathbf{x},\frac{\Delta t}{\epsilon}\right)-\mathcal{P}\left(\nabla_{\perp}^{2}T^{n+1}+S^{n+1};\mathbf{x},\frac{2\Delta t}{\epsilon}\right)\right].
\end{split}
\label{eq:bdf2-linear-system}
\end{align}
We solve these linear equations iteratively with a GMRES Krylov solver
\citep{saad1986gmres}, preconditioned as explained earlier in this
study. The propagators are evaluated at every GMRES iteration (see
below). While our original transport equation is self-adjoint for
certain boundary conditions, the numerical implementation of the semi-Lagrangian
forms in Eqs. (\ref{eq:bdf1-linear-system}) and (\ref{eq:bdf2-linear-system})
cannot be guaranteed to be strictly symmetric, and therefore we choose
GMRES instead of CG as our Krylov solver. Unless otherwise specified,
for convergence we enforce a relative decrease of the linear GMRES
residual of $10^{-4}$. The size of the Krylov subspace in GMRES is
kept large enough to accomodate the iteration count in the results
below without restarting.

The Lagrangian $\mathcal{P}$ propagators in Eqs. (\ref{eq:bdf1-linear-system})
and (\ref{eq:bdf2-linear-system}) interpolate $\nabla_{\perp}^{2}T^{n+1}$
along magnetic field lines, and this results in a strong nonlocal
coupling that makes forming the corresponding system matrix impractical.
Instead, we use a matrix-free implementation of GMRES, in which the
linear operators are evaluated (and therefore a Lagrangian step is
performed) every time GMRES requires a matrix-vector product. This
makes the iteration practical (because a matrix is not explicitly
built), but expensive, particularly for very small $\epsilon$, thereby
putting a premium on effective preconditioning.

\subsection{Numerical integration of Lagrangian integrals}

The evaluation of the Lagrangian integrals requires integration of
the magnetic field lines (and associated propagators) originating
at every point of the mesh, which is the most computationally expensive
part of the algorithm. The numerical implementation of these field-line
integrals reuses the infrastructure developed in previous work \citep{chacon2014asymptotic},
and follows the setup outlined in the reference closely. The kernel
integrals are reformulated as ordinary differential equations (ODEs),
and solved in conjunction with the magnetic field line equation (\ref{trace_mag})
with the high-order ODE integration package ODEPACK \citep{hindmarsh1983odepack}.
The absolute and relative tolerances of the ODE solver and the integral
error estimates are kept very tight ($10^{-14}$) to prevent impacting
solution accuracy and solver performance.

In practice, $T^{n+1}$, $T^{n}$, $S^{n+1}$, and the magnetic field
(even when analytical formulas exist) in Eqs. (\ref{eq:bdf1-linear-system})
and (\ref{eq:bdf2-linear-system}) are provided on a computational
grid, so the Lagrangian integrals in the operators $\mathcal{G}$
and $\mathcal{P}$ require the reconstruction (by interpolation) of
these discrete fields over the whole domain to evaluate them at arbitrary
points along magnetic field orbits. This is done in this study with
global, arbitrary-order splines, but we have also implemented and
tested second-order B-spline-based positivity-preserving local-stencil
interpolations (more suitable for massively parallel applications,
but less accurate; see Ref. \citep{chacon2022asymptotic} for details).

\subsection{Perpendicular transport}

The perpendicular transport operator $\mathbb{B}$ inside the formal
source $S_{*}$ is discretized using either second-order or fourth-order
conservative finite differences in the evaluation of the GMRES linear
residual, and with second-order finite differences in the preconditioner.
For a spatially constant perpendicular transport coefficient $\chi_{\perp}$,
one may reformulate equation (\ref{diffusion_eq}) as: 
\begin{equation}
\partial_{t}T-\left(\frac{1}{\epsilon}-1\right)\nabla_{\parallel}^{2}T=\nabla^{2}T+S\equiv S_{*}.
\end{equation}
Thus, replacing $1/\epsilon$ with $1/\epsilon-1$ and substituting
perpendicular Laplacian $\nabla_{\perp}^{2}$ with isotropic one $\nabla^{2}$,
yields the same system. The analysis and method formulation in previous
sections stays intact, but now: 
\begin{equation}
\alpha\rightarrow\Delta tk_{\parallel}^{2}\left(\frac{1}{\epsilon}-1\right),\quad\beta\rightarrow\Delta tk^{2},\quad\mathbb{B}\rightarrow-\Delta t\nabla^{2},\label{beta_def}
\end{equation}
This formulation is more desirable because a linear, positivity-preserving
discretization of the isotropic Laplacian exists, and it is better
behaved solverwise. If such a reformulation is not possible, a positivity-preserving
treatment of the perpendicular transport operator is still possible
\citep{dutoit2018positivity}, albeit the discretization is rendered
nonlinear.

\section{Numerical tests}

\label{sec:ntests} In this section, we demonstrate the spatio-temporal
accuracy properties of the implicit semi-Lagrangian scheme and the
performance of the proposed preconditioner with several challenging
tests. The first accuracy test (Section~\ref{sec:2zone}) is intended
to isolate the accuracy benefits of the scheme in the presence of
boundary layers (e.g., magnetic island separatrices), and employs
an analytically solvable problem for this purpose. The test confirms
marked accuracy improvements over the earlier operator-split formulation
\citep{chacon2014asymptotic}, which was particularly challenged by
such configurations. The second accuracy test (Section~\ref{sec:acc-complex-B})
focuses on demonstrating the accuracy of the scheme in complex magnetic
field topologies, featuring magnetic islands, with extreme anisotropies
($\chi_{\parallel}/\chi_{\perp}=10^{10}$). Finally, in Section \ref{sec:convergence_tests}
we demonstrate the performance of the scheme for both straight and
curved magnetic fields, and confirm the results of the convergence
analysis in Sec. \ref{sec:prec}.

\subsection{Accuracy test in the presence of boundary layers}

\label{sec:2zone} %%%%%%%%%%%%%%%%%%%%%%%%%%%%%%%%%%%%%
%%%%%  two zone problem         %%%%%
%%%%%%%%%%%%%%%%%%%%%%%%%%%%%%%%%%%%%
The two-zone problem is a proxy for a magnetic island problem, which
retains essential physics of magnetic islands, but features an analytical
solution. The motivation for this test is two-fold. First, magnetic
islands pose challenges for conventional numerical schemes (such as
distinct topological regions), which the new semi-Lagrangian algorithm
is designed to overcome. Therefore, a problem with an exact analytical
solution but with similar challenges is perfect for testing the new
algorithm. Second, the solution of the two-zone problem (as for islands)
has a boundary layer of length determined by the anisotropy ratio
($\delta\sim\sqrt{\epsilon}$). The previous operator-split formulation
\citep{chacon2014asymptotic} had an accuracy-based time-step limitation
($\Delta t<\sqrt{\epsilon}$), which is problematic in the presence
of boundary layers. As we will show, the implicit approach features
no such time-step constraint, and therefore the test clearly demonstrates
the advantage of the new formulation.

\begin{figure}[t]
\centering \includegraphics[width=0.8\columnwidth]{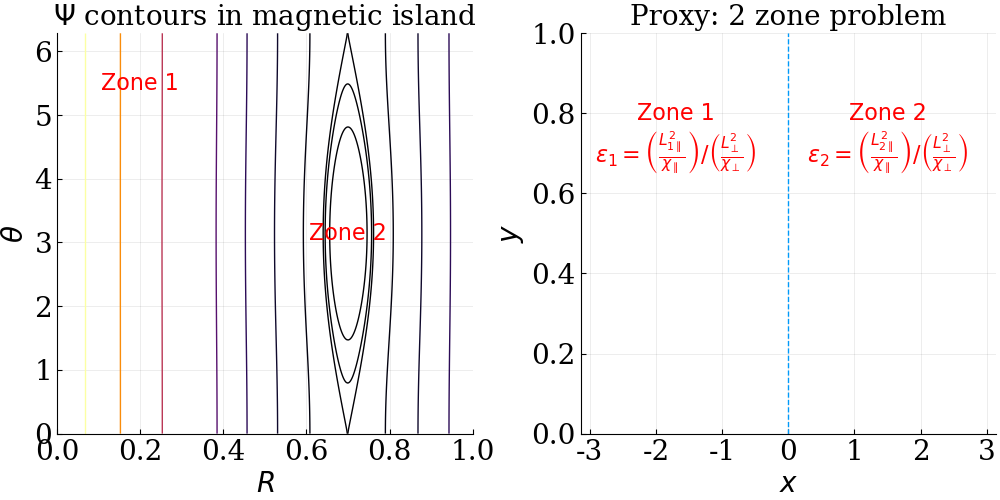}
\caption{Schematic representation of the unrolling magnetic island into a two-zone
configuration. The left figure shows a typical magnetic field topology
in cylindrical/toroidal geometry (poloidal angle vs radius). The right
figure shows the Cartesian unrolling into a two-zone problem.}
\label{fig:island_scheme}
\end{figure}

The main idea to emulate the magnetic island is to have two distinct
zones with different anisotropy ratio, i.e., 
\begin{equation}
\epsilon=\epsilon(x)=\begin{cases}
 & \epsilon_{1}\text{, for }x\in[-\pi,0],\\
 & \epsilon_{2}\text{, for }x\in(0,\pi],
\end{cases}\label{epsilon}
\end{equation}
with $\epsilon_{1}\neq\epsilon_{2}$ (we choose $\epsilon_{1}\gg\epsilon_{2}$).
The jump in $\epsilon$ mimics the sudden change in magnetic field
topology (i.e., field line length), inside and outside the magnetic
island, as schematically shown in Figure~\ref{fig:island_scheme}.
This is the case because the dimensionless anisotropy ratio 
\[
\epsilon=\frac{\chi_{\perp}}{\chi_{\parallel}}\left(\frac{L_{\parallel}}{L_{\perp}}\right)^{2},
\]
depends on the magnetic field line length, $L_{\parallel}$.

The two-zone test is two dimensional, so $T=T(t,x,y)$, $x\in[-\pi,\pi]$,
$y\in[0,1]$, with homogeneous Dirichlet boundary conditions in $x$
and periodic in $y$. In this problem, we assume zero initial conditions
$T(0,x,y)=0$, and the temperature profile evolves to establish equilibrium
due to a localized external heat source: 
\begin{equation}
S=S(x,y)=\begin{cases}
 & -\sin(x)\sin(k_{y}y)\text{, for }x\in[-\pi,0],\\
 & 0\text{, for }x\in(0,\pi],
\end{cases}\label{2z:source}
\end{equation}
where $k_{y}=2\pi$. The magnetic field is straight and uniform along
the $y$ axis. 

\begin{figure}[t]
\centering \includegraphics[width=0.5\columnwidth]{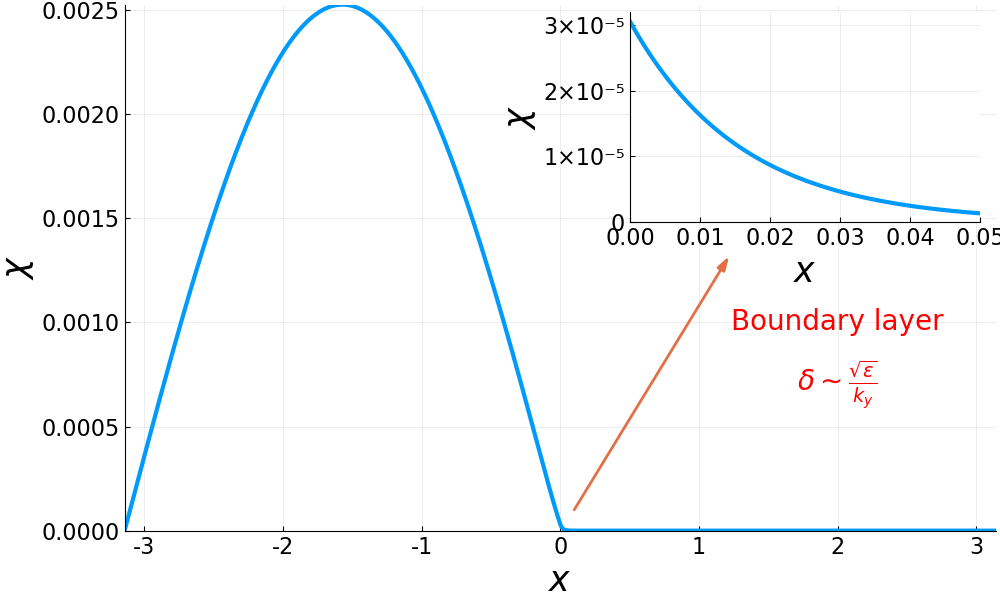}
\caption{\label{fig:blayer} Typical steady state solution (\ref{chi}) and
illustration of the boundary layer.}
\end{figure}

A linear analysis (see \ref{2z:ss}) shows that the steady-state solution
to the problem for $k_{y}\neq0$ is 
\begin{equation}
T_{s}(x,y)=\chi(x)\sin(k_{y}y),\label{2z:ss_sol}
\end{equation}
with 
\begin{equation}
\chi(x)=\begin{cases}
 & -\frac{1}{1+r_{1}^{2}}\sin(x)+A\frac{\sinh\left(r_{1}(x+\pi)\right)}{\sinh\left(r_{1}\pi\right)}\text{, for }x\in[-\pi,0],\\
 & A\frac{\sinh\left(r_{2}(\pi-x)\right)}{\sinh\left(r_{2}\pi\right)}\text{, for }x\in(0,\pi],
\end{cases}\label{chi}
\end{equation}
where $r_{1,2}=k_{y}/\sqrt{\epsilon_{1,2}}$ and 
\[
A=\frac{1}{1+r_{1}^{2}}\frac{1}{r_{1}\coth\left(\pi r_{1}\right)+r_{2}\coth\left(\pi r_{2}\right)}.
\]
The typical shape of the solution along $x$ is plotted in Figure~\ref{fig:blayer}.
This figure illustrates that the solution mimics the heat source (\ref{2z:source})
in zone 1, decaying immediately at the beginning of zone 2. This transition
region is a boundary layer. Taylor-expanding equation (\ref{chi})
shows that the boundary layer width can be estimated as 
\begin{equation}
\delta\approx\frac{1}{r_{2}}=\frac{\sqrt{\epsilon_{2}}}{k_{y}}.\label{blayer_width}
\end{equation}

The derivation of the full time-dependent solution for the two-zone
problem is described in \ref{2z:time}. The elliptic nature of the
diffusion equation makes the temperature profile relax to the steady-state
solution (\ref{2z:ss_sol}), 
\begin{equation}
T(t,x,y)=T_{s}(x,y)+\sum_{n=1}^{\infty}C_{n}e^{-\gamma_{n}t}h_{n}(x,y),
\end{equation}
where $\gamma_{n}>0$ are real damping rates that can be found from
the transcendental dispersion equation in \ref{2z:time} {[}Eq. (\ref{disp}){]},
$h_{n}$ are the eigenfunction of the homogeneous diffusion equation
found in \ref{2z:time}, and $C_{n}=-\langle T_{s},h_{n}\rangle/\langle h_{n},h_{n}\rangle$,
with the appropriate integral scalar product $\langle f,g\rangle=\int fgdxdy$,
according to the classical Sturm-Liouville theory.

\begin{figure}[htb]
\centering \includegraphics[width=0.4\columnwidth]{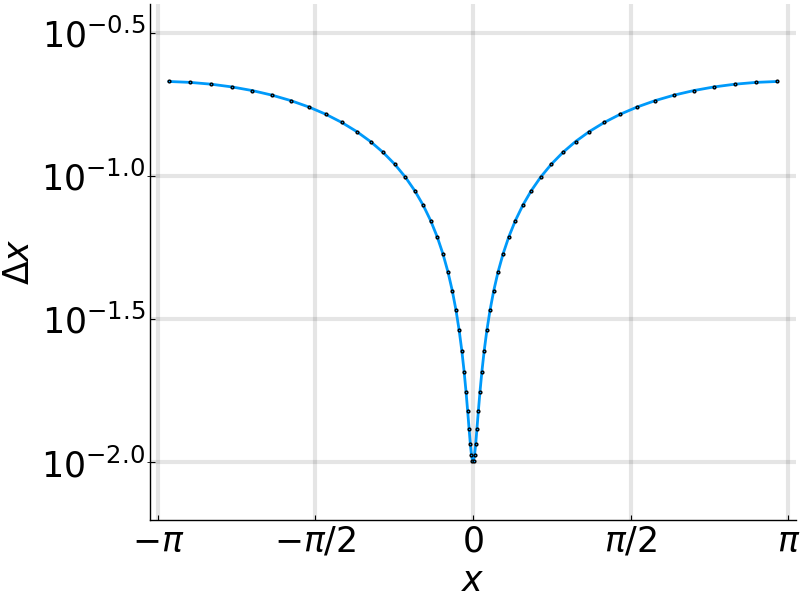} \caption{Spatial variation of the mesh size in $x$ direction in two-zone problem.}
\label{fig:sstate:mesh}
\end{figure}

\begin{figure}[h!tbp]
\centering %    \includegraphics[width=0.6\linewidth]{sstate2}
\begin{subfigure}[t]{3in} \includegraphics[width=1\linewidth]{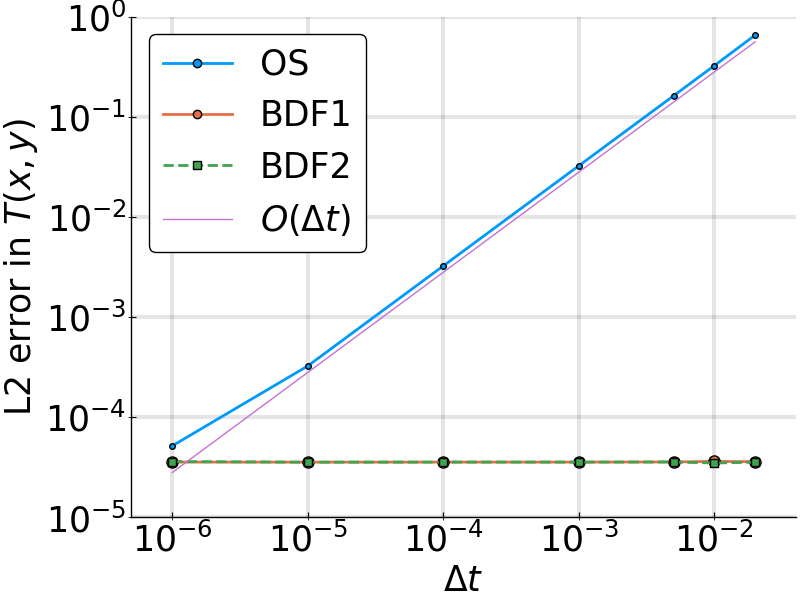}
\caption{with $\epsilon_{1}=10^{-1},\epsilon_{2}=10^{-2}$}
\label{fig:sstate2} \end{subfigure} \begin{subfigure}[t]{3in}
\includegraphics[width=1\linewidth]{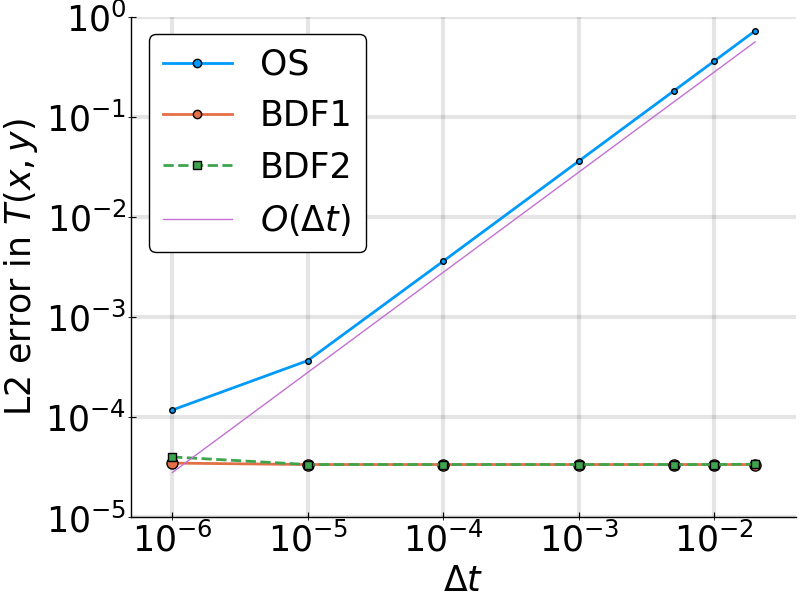} \caption{with $\epsilon_{1}=10^{-3},\epsilon_{2}=10^{-4}$}
\label{fig:sstate4} \end{subfigure} \caption{Error in steady-state solution for three different time discretizations:
BDF1, BDF2, and operator splitting in two-zone problem versus time
step.}
\label{fig:sstate}
\end{figure}

In the first numerical test, we verify that the implicit time discretization
can recover the steady-state solution up to spatial discretization
errors regardless of the time step. We evolve the simulation for a
sufficiently long time $t_{final}=0.04$ to recover the steady-state
solution, and vary the time step. We test three time discretization
algorithms: BDF1, BDF2, and operator splitting (OP) \citep{chacon2014asymptotic}.
We fix the spatial resolution, with the number of points in $x$ to
be $N_{x}=64$ and in $y$, $N_{y}=32$. Note that the spatial grid
is uniform in $y$, but it is packed in $x$ (i.e., tensor product
mesh) around the boundary layer ($x=0$), so the boundary layer is
always well resolved. We will relax this later. The actual spatial
variation of the mesh size in $x$ direction is shown in Figure~\ref{fig:sstate:mesh}.
Two anisotropy values are chosen for this test, $\epsilon_{1}=10^{-1},\epsilon_{2}=10^{-2}$
and $\epsilon_{1}=10^{-3},\epsilon_{2}=10^{-4}$. We measure the $L_{2}$
error between numerical and analytical steady-state solutions and
the results are shown in Figure~\ref{fig:sstate}. It is evident
form the graph that the operator-split method is first order in time.
The steady-state errors in the implicit temporal schemes, however,
do not depend on the time step, and the deviation from the analytical
solution is only due to spatial error. 

\begin{figure}[t]
\centering \includegraphics[width=0.4\columnwidth]{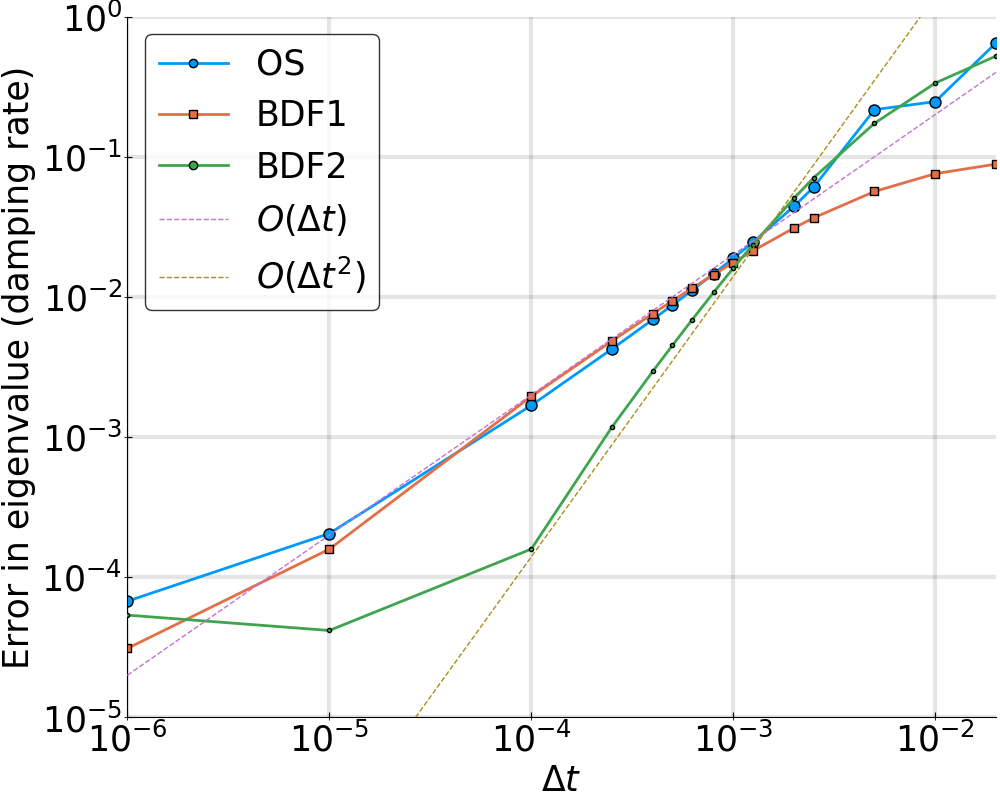} \caption{Relative temporal discretization error in the damping rate $\gamma_{1}$
versus time-step of the first eigenfunction $h_{1}$ for three different
time discretizations: BDF1, BDF2, and OS in the two-zone problem with
$\epsilon_{1}=10^{-1},\epsilon_{2}=10^{-2}$.}
\label{fig:time}
\end{figure}

Next, we test the time discretization error with the full time-dependent
solution. For this test, we modify the initial condition as: 
\begin{equation}
T(0,x,y)=T_{s}(x,y)+h_{1}(x,y),
\end{equation}
where $h_{1}(x,y)$ is the slowest decaying eigenmode, so the temperature
profile evolves exactly as 
\begin{equation}
T(t,x,y)=T_{s}(x,y)+e^{-\gamma_{1}t}h_{1}(x,y).
\end{equation}
Here, $\gamma_{1}$ is determined by (\ref{gamma1}) and $h_{1}$
by (\ref{2z:full_hom_sol}). In this test, we choose the anisotropy
ratios $\epsilon_{1}=10^{-1},\epsilon_{2}=10^{-2}$, with corresponding
theoretical damping rate $\gamma_{1}^{theory}=395.774$. We measure
the experimental damping rate $\gamma_{1}$ by a linear regression
of the expression 
\[
\log\left(\left\Vert T-T_{s}\right\Vert _{L_{2}}\right)=-\gamma_{1}t+\log\left(\left\Vert h_{1}\right\Vert _{L_{2}}\right),
\]
where $\left\Vert \cdot\right\Vert _{L_{2}}$ is an $L_{2}$ norm.
The results of the relative error in the damping rate are shown in
Figure~\ref{fig:time}, where we can see that BDF1 and OS methods
are first-order in time while BDF2 is second-order. 

\begin{figure}[t]
\centering \includegraphics[width=0.4\columnwidth]{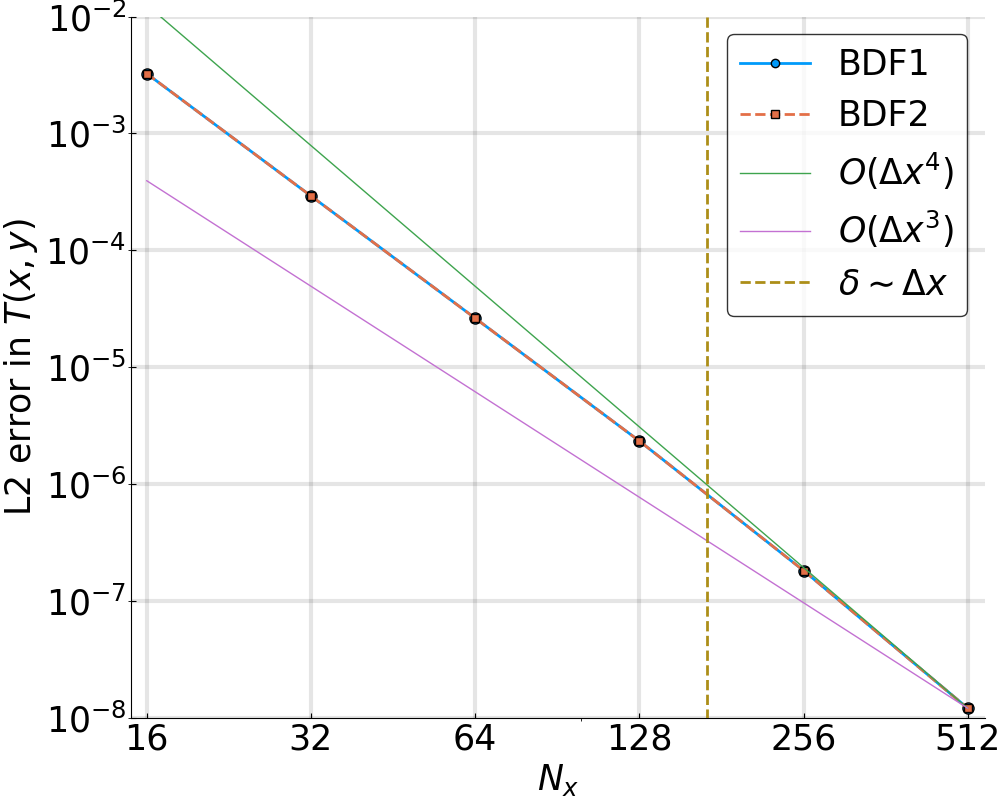}
\caption{Spatial discretization error in the steady-state solution for BDF1
and BDF2 temporal discretizations in the two-zone problem with $\epsilon_{1}=10^{-1},\epsilon_{2}=10^{-2}$
versus the number of spatial points. The vertical dashed line indicates
the width of the boundary layer for this configuration.}
\label{fig:packing}
\end{figure}

Next, we test the numerical error due to the spatial discretization
in the steady-state solution. We consider a uniform grid both in $x$
and $y$, and refine the number of spatial points $N_{x}=N_{y}$ (other
parameters are left unchanged from the previous test). The results
of the $L_{2}$ error are shown in Figure~\ref{fig:packing}. We
can see that, when $\Delta x$ is smaller than the boundary layer
width Eq. (\ref{blayer_width}), BDF1 and BDF2 are both fourth-order
accurate spatially, as expected. However, there is a slight order
reduction (less than one order), when the boundary layer is not resolved.
The steady-state spatial accuracy is independent of the temporal discretization,
also expected.

\subsection{Spatial accuracy test with complex magnetic field topologies and
extreme anisotropies ($\chi_{\parallel}/\chi_{\perp}=10^{10}$)}

\label{sec:acc-complex-B}

\begin{figure}[h!tbp]
\centering{}\includegraphics[width=0.45\columnwidth]{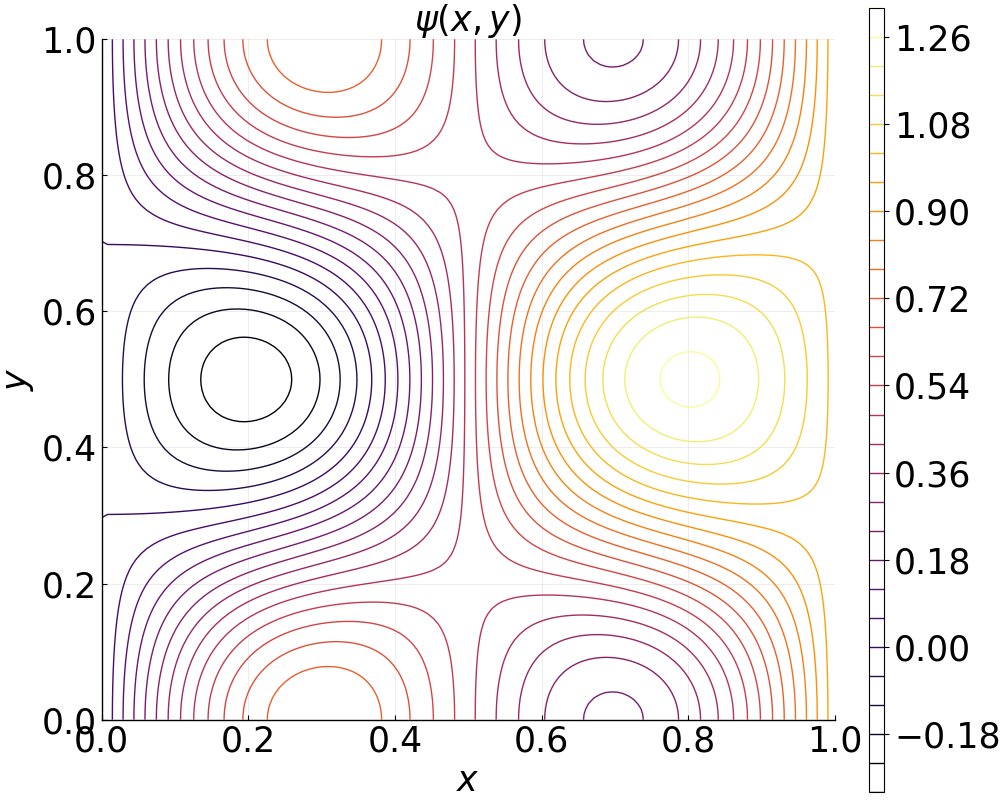}~\includegraphics[width=0.4\columnwidth]{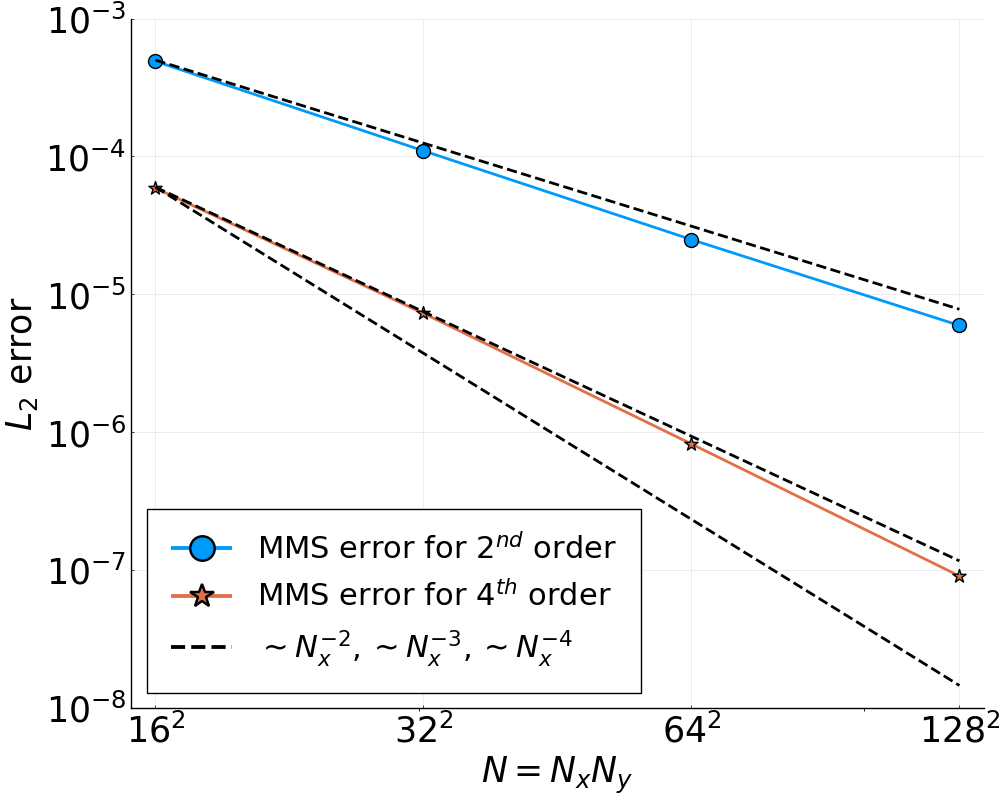}
\caption{\label{fig:complex_acc_eps=00003D1e-10} Left: Poloidal flux for manufactured
solution test with extreme anisotropy. Right: Scaling of spatial error
in steady state with respect to the manufactured solution $T_{\infty}$
for $\chi_{\parallel}/\chi_{\perp}=10^{10}$, demonstrating desing
accuracy for the second-order discretization, and convergence between
third and fourth order for our fourth-order implementation. Error
is measured with an $L_{2}$-norm with respect to the analytical steady
state, and is normalized to the maximum temperature value.}
\end{figure}

In order to test the spatial accuracy of the semi-Lagrangian formulation
with extreme anisotropies, we consider a 2D domain with $L_{x}=L_{y}=1$
and homogeneous Dirichlet boundary conditions in $x$ and periodic
boundary conditions in $y$. We consider a steady-state manufactured
solution of the form $T_{\infty}(x,y)=\psi(x,y)$, with $\psi$ the
poloidal flux given by 
\begin{equation}
\psi=x+\delta\sin(2\pi x)\cos(2\pi y),\label{eq:poloidal_flux}
\end{equation}
with $\delta=1/2$ (for which $\psi$ features several magnetic islands;
see Figure~\ref{fig:complex_acc_eps=00003D1e-10}-left). The magnetic
field is given by $\mathbf{B}=\mathbf{z}\times\nabla\Psi+B_{z}\mathbf{z}$,
with $B_{z}=1$. With these choices, $T_{\infty}(x,y)$ is in the
kernel of the parallel transport operator, i.e., $\nabla_{\parallel}^{2}T_{\infty}=0$.
This is similar to other tests proposed in the literature \citep{sovinec2004nonlinear},
and is designed to allow a clean measurement of numerical error pollution
by the stiff parallel transport dynamics. The source $S(\mathbf{x})$
in Equation (\ref{sec:acc-complex-B}) that drives the temperature
field to this steady-state solution is found as:
\[
S(x,y)=-\nabla_{\perp}^{2}T_{\infty}=-\nabla^{2}\psi=4\pi^{2}\sin(2\pi x)\cos(2\pi y).
\]
The results of the convergence study in Figure \ref{fig:complex_acc_eps=00003D1e-10}-right
demonstrate spatial convergence close to or at design accuracy for
both second- and fourth-order discretizations, with relative errors
small even for very coarse meshes. This confirms the ability of the
scheme to capture null-space components of the solution accurately,
without pollution from the parallel-transport dynamics.

\subsection{Performance tests}

\label{sec:convergence_tests}We assess next the performance of the
preconditioner proposed in Section~\ref{sec:prec} for both trivial
($\delta=0.1$) and complex ($\delta=0.5$) magnetic field topologies
with extreme anisotropy, $\chi_{\parallel}/\chi_{\perp}=10^{10}$.
For this test, we use the second-order discretization of the isotropic
Laplacian operator.

A convergence study for a GMRES tolerance of $\epsilon_{r}=10^{-3}$
is presented in Table \ref{tab:Convergence-study-6} for the same
setup as in Section \ref{sec:acc-complex-B} for two values of the
parameter $\delta$ (0.1 and 0.5), and for various mesh sizes and
timesteps. For $\delta=0.1$, the magnetic field is simply connected,
and no boundary layers exist (not shown). The Table demonstrates no
dependence of the convergence rate with mesh refinement, and very
weak dependence on the timestep, as expected from the discussion in
Sec. \ref{sec:prec}.

For $\delta=0.5$, the magnetic field topology features islands, and
therefore boundary layers. Accordingly, off-null-space components
will be present in the solution to enforce continuity of the temperature
field across these boundary layers. This manifests in the results
in the Table, where sensitivity to both resolution and timestep can
be observed. However, there is a marked transition in performance
between $\Delta t=0.01$ and $\Delta t=0.1$. Below $\Delta t=0.01$,
the convergence rate of the scales as $\sim(N\Delta t)^{1/4}$. Above
$\Delta t=0.1$, the convergence rate becomes largely independent
of mesh refinement and improves with larger $\Delta t$ (both a consequence
of improved eigenvalue clustering). For this problem, the $\beta_{\mathrm{min}}=4\pi^{2}\chi_{\perp}\Delta t/L^{2}>1$
threshold advanced in Sec. \ref{sec:prec} corresponds to $\Delta t>L^{2}/(4\pi^{2}\chi_{\perp})\approx0.025$,
which fully explains the transition in convergence behavior. When
the residual is analyzed during the course of the iteration for $\Delta t<0.025$,
the error is seen to concentrate around the island separatrices (not
shown), suggesting that convergence is hampered by the presence of
off-null-space components. The unpreconditioned case recovers the
standard dependence of the Krylov convergence rate with mesh size
and timestep, $\sqrt{N\Delta t}$ for all timesteps considered, demonstrating
the effectiveness of the preconditioner in decreasing the iteration
count.
\begin{table}
\caption{\label{tab:Convergence-study-6}\textcolor{black}{Convergence study
for various mesh sizes and timesteps for $\chi_{\parallel}/\chi_{\perp}=10^{10}$
and $\delta=0.1$, 0.5 with the second-order discretization. Timesteps
considered $\Delta t=10^{-5},0.001,0.01,0.1,1.0$. Note that the first
timestep is the accuracy limit for the operator-split algorithm in
Ref. \citep{chacon2014asymptotic} ($\Delta t=\sqrt{\frac{\chi_{\perp}}{\chi_{\parallel}}}$).
The numbers indicated by ``id'' indicate unpreconditioned GMRES
iterations.}}

\centering{}%
\begin{tabular}{|c|c|c|c|c|c|c|c|c|c|c|}
\hline 
 & \multicolumn{2}{c|}{$\Delta t=10^{-5}$} & \multicolumn{2}{c|}{$\Delta t=0.001$} & \multicolumn{2}{c|}{$\Delta t=0.01$} & \multicolumn{2}{c|}{$\Delta t=0.1$} & \multicolumn{2}{c|}{$\Delta t=1.0$}\tabularnewline
\hline 
Mesh/$\delta$ & $0.1$ & $0.5$ & $0.1$ & $0.5$ & $0.1$ & $0.5$ & $0.1$ & $0.5$ & $0.1$ & $0.5$\tabularnewline
\hline 
\hline 
$32\times32$ & 1 & 2 & 4 & 9 (id=9) & 8 & 22 (id=24) & 10 & 20 (id=51) & 14 & 16\tabularnewline
\hline 
$64\times64$ & 1 & 2 & 4 & 15 (id=16) & 8 & 36 (id=45) & 6 & 22 (id=83) & 10 & 13\tabularnewline
\hline 
$128\times128$ & 1 & 4 & 5 & 23 (id=30) & 8 & 46 (id=84) & 5 & 21 (id=111) & 4 & 17\tabularnewline
\hline 
$256\times256$ & 1 & 5 & 5 & 37 (id=62) & 8 & 63 (id=196) & 5 & 19 (id=269) & 3 & 14\tabularnewline
\hline 
\end{tabular}
\end{table}

\section{Discussion}

\label{sec:dis} We have developed and implemented a new preconditioned
implicit solver for the semi-Lagrangian asymptotically preserving
scheme for the strongly anisotropic heat transport equation proposed
in \citep{chacon2014asymptotic}. The implicit algorithm is demonstrated
to be much more accurate than the operator-split one proposed in the
reference, both for steady-state solutions as well as solutions with
boundary layers.

GMRES solver performance is independent of the anisotropy ratio, which
is critical. We have proposed a simple preconditioner that results
in the number of GMRES iterations $N_{i}$ scaling as $(N\Delta t)^{1/4}$,
with $N$ the total number of mesh points. Since our equations are
normalized to the perpendicular transport time scale, $\tau_{\perp}=L_{\perp}^{2}/\chi_{\perp}$,
and $N=(L/\Delta)^{2}$, with $\Delta$ the mesh spacing, one can
rewrite this result as $N_{i}\sim\left(\frac{\Delta t}{\tau_{\perp}}\right)^{1/4}\sqrt{\frac{L_{\perp}}{\Delta}}$.

While in principle the method does not scale optimally under mesh
and time-step refinement, it is still instructive to compare its computational
performance vs the operator-split one of Ref. \citep{chacon2014asymptotic}.
We can estimate the computational cost of the operator splitting method
as: 
\begin{equation}
\mbox{CPU time}\sim\frac{T^{final}}{\Delta t_{OS}}\sim\frac{T^{final}}{\tau_{\perp}\sqrt{\chi_{\perp}/\chi_{\parallel}}},
\end{equation}
where the time step restriction $\Delta t_{OS}\sim\tau_{\perp}\sqrt{\chi_{\perp}/\chi_{\parallel}}$
is needed to avoid $O(1)$ errors at boundary layers (e.g., at magnetic
island separatrices). %Here, following \cite{chacon2014asymptotic}, $\tau_\perp = L_\perp^2/\chi_\perp$, with $L_\perp$ the characteristic perpendicular diffusion length scale.
At the same time, the cost of the implicit method is:
\begin{equation}
\mbox{CPU time}\sim\frac{T^{final}N_{i}}{\Delta t}.
\end{equation}
The speed-up can therefore be estimated as: 
\begin{equation}
\mbox{speedup}=\frac{\left(\mbox{CPU time}\right)_{OS}}{\left(\mbox{CPU time}\right)_{impl}}\sim\sqrt{\frac{\Delta}{L_{\perp}}}\left(\frac{\Delta t}{\tau_{\perp}}\right)^{3/4}\sqrt{\frac{\chi_{\parallel}}{\chi_{\perp}}}.
\end{equation}
This expression implies that the implicit method favors large time
steps and high anisotropy, and in this regime it is significantly
more efficient than the operator-split approach.

Finally, it is worth noting that, in many practical applications of
interest (and in particular for magnetic thermonuclear fusion) for
which $\chi_{\perp}$ is very small, time and mesh resolution requirements
compatible with other physical processes (e.g. magnetohydrodynamics)
satisfy $\Delta t\sim\tau_{\perp}$ and $\Delta\sim L_{\perp}$, thus
leading to $N_{i}\sim\mathcal{O}(1)$ and making the algorithm competitive
in practice.

\section{Conclusion}

\label{sec:con} In this study, we have investigated the merits and
capabilities of a new fully implicit, asymptotic-preserving, semi-Lagrangian
algorithm for the time-dependent anisotropic heat-transport equation.
The new method is an implicit extension of the operator-split approach
proposed in \citep{chacon2014asymptotic}. The integro-differential
formulation of the parallel transport operator is the key element
to ensure asymptotically preserving properties in the limit of arbitrary
anisotropy, thereby avoiding numerical pollution. The perpendicular
transport is treated as a formal source to the purely parallel transport
equation, thus constituting an implicit integro-differential equation.
The resulting linear system is inverted with GMRES, and preconditioned
to be compact. For complex magnetic-field topologies and below a timestep
size threshold, the preconditioner results in Krylov convergence in
$N_{i}\sim(N\Delta t)^{1/4}$ iterations, i.e., scaling very weakly
with mesh refinement and time-step size and independently of the anisotropy
ratio $\epsilon$. Above the timestep threshold, convergence is essentially
mesh- and timestep-independent. Additionally, the new implicit formulation
is unconditionally stable and preserves the positivity of the solution
for physical initial conditions.

Similarly to the operator-split formulation, the implicit version
can handle complicated magnetic field topologies and very large anisotropy
ratios without introducing numerical pollution. However, unlike the
operator-split approach, its implicit nature ensures accurate steady
states regardless of $\Delta t$, and can be readily made higher-order
temporally (e.g., BDF2). The implicit formulation favors large time
steps and high anisotropy, with speedups vs. the operator-split approach
$\sim\Delta t^{3/4}\sqrt{\chi_{\parallel}/\chi_{\perp}}$.

The proposed approach features a number of simplifications that make
it unsuitable for some class of problems. The main two limitations
are the approximation $\nabla\cdot\mathbf{b}\approx0$ (for which
we have devised a solution that will be documented in a future publication)
and the assumption of constant parallel conductivity along field lines
(arbitrary variations of $\chi_{\perp}$ and variation of $\chi_{\parallel}$
across field lines can be handled with the current formulation of
the method). The extension of the method to remove the latter limitation
will be considered in the future.

\section*{Acknowledgements}

This work was supported in parts by Triad National Security, LLC under
contract 89233218CNA000001 and DOE Office of Applied Scientific Computing
Research (ASCR). The research used computing resources provided by
the Los Alamos National Laboratory Institutional Computing Program.
The authors are grateful for useful and informative discussions with
C. T. Kelley.

\appendix

\section{Steady-state solution to the two-zone problem}

\label{2z:ss} In order to derive the steady-state solution (\ref{2z:ss_sol}),
we look for a solution of the form $T=\chi(x)\sin(k_{y}y)$, such
that: 
\begin{align}
 & \chi''_{1}-\frac{k_{y}^{2}}{\epsilon_{1}}\chi_{1}=\sin(x),\\
 & \chi''_{2}-\frac{k_{y}^{2}}{\epsilon_{2}}\chi_{2}=0,
\end{align}
where indices $1$ and $2$ correspond to zone 1 ($x\in[-\pi,0]$)
and 2 ($x\in[0,\pi]$), respectively. The solution which satisfies
the boundary conditions 
\[
\chi_{1}(-\pi)=\chi_{2}(\pi)=0,\quad\chi_{1}(0)=\chi_{2}(0),\quad\chi_{1}'(0)=\chi_{2}'(0),
\]
is: 
\begin{align}
 & \chi_{1}(x)=-\frac{1}{1+r_{1}^{2}}\sin(x)+A\frac{\sinh\left(r_{1}(x+\pi)\right)}{\sinh\left(r_{1}\pi\right)},\\
 & \chi_{2}(x)=A\frac{\sinh\left(r_{2}(\pi-x)\right)}{\sinh\left(r_{2}\pi\right)},
\end{align}
where $r_{1,2}=k_{y}/\sqrt{\epsilon_{1,2}}$ and 
\[
A=\frac{1}{1+r_{1}^{2}}\frac{1}{r_{1}\coth\left(\pi r_{1}\right)+r_{2}\coth\left(\pi r_{2}\right)},
\]
which is (\ref{chi}). %%%%%%%%%%%%%%%%%%%%%%%%%%%%%%
%%%%  Damping rates      %%%%%
%%%%%%%%%%%%%%%%%%%%%%%%%%%%%%

\section{Time-dependent solution of the two-zone problem}

\label{2z:time} In this section, we extend the analysis of \ref{2z:ss}
to include the time evolution of the solution. Since the steady-state
solution is already known and the equation (\ref{diffusion_eq}) is
linear, it is sufficient to solve the homogeneous equation 
\begin{equation}
\partial_{t}T-\frac{1}{\epsilon}\partial_{y}^{2}T-\partial_{x}^{2}T=0,\label{hom}
\end{equation}
with $\epsilon$, boundary conditions defined in Sections \ref{sec:2zone},
and initial condition $T(0,x,y)=-T_{s}(x,y)$ where $T_{s}$ is the
steady-state solution from \ref{2z:ss} (note that the initial condition
of the initial problem is zero).

Separation of variables $T=\tau(t)X(x)\sin(k_{y}y)$ (with $k_{y}\neq0$)
leads to 
\begin{equation}
\tau(t)=\tau(0)e^{-\gamma t},
\end{equation}
where eigenvalues $\gamma$ (damping rates) and eigenfunctions $X(x)$
are determined from the boundary value problem (note the discontinuity
in $\epsilon$) 
\begin{align}
 & X''-\left(\frac{k_{y}^{2}}{\epsilon}-\gamma\right)X=0,\\
 & X(-\pi)=X(\pi)=0.
\end{align}
After some algebra, the boundary value problem solution yield 
\begin{equation}
X(x)=\begin{cases}
 & \frac{\sin(\sigma_{1}(\pi+x))}{\sin(\pi\sigma_{1})}\text{, for }x\le0,\\
 & \frac{\sinh(\lambda_{2}(\pi-x))}{\sinh(\pi\lambda_{2})}\text{, for }x\ge0,
\end{cases}
\end{equation}
with $\sigma_{1}=\sqrt{\gamma-r_{1}^{2}}$, and $\lambda_{2}=\sqrt{r_{2}^{2}-\gamma}$
(we also define $\lambda_{1}=\sqrt{r_{1}^{2}-\gamma}$, and $\sigma_{2}=\sqrt{\gamma-r_{2}^{2}}$
) where $\gamma$ is a solution to the dispersion equation 
\begin{equation}
\frac{\tan(\pi\sigma_{1})}{\pi\sigma_{1}}+\frac{\tanh(\pi\lambda_{2})}{\pi\lambda_{2}}=0.\label{disp}
\end{equation}
Using the eigenmodes obtained from the boundary value problem, we
derive the solution of the homogeneous system (\ref{hom}) 
\begin{equation}
T=\sin(k_{y}y)\sum_{n}C_{n}e^{-\gamma_{n}t}\begin{cases}
 & \frac{\sin(\sigma_{1}^{n}(\pi+x))}{\sin(\pi\sigma_{1}^{n})}\text{, for }x\le0,\\
 & \frac{\sinh(\lambda_{2}^{n}(\pi-x))}{\sinh(\pi\lambda_{2}^{n})}\text{, for }x\ge0,
\end{cases}\label{2z:full_hom_sol}
\end{equation}
where summation is performed over all eigenmodes and corresponding
eigenvalues $\gamma^{n}$ (and $\sigma_{1}^{n}$, $\lambda_{2}^{n}$)
found from the dispersion equation (\ref{disp}). The coefficient
$C_{n}$ should be computed from the initial condition expanded in
a basis of eigenfunctions found from the boundary value problem.

It is impossible to solve the dispersion equation (\ref{disp}) analytically.
However, it is possible to estimate the smallest growth rate for the
parameters of interest. First, we notice that there are no roots for
$\mbox{Im}[\lambda_{1}]=\mbox{Im}[\lambda_{2}]=0$, thus the smallest
$\gamma$ will appear in the regime 
\begin{equation}
\frac{k_{y}^{2}}{\epsilon_{1}}<\gamma<\frac{k_{y}^{2}}{\epsilon_{2}},
\end{equation}
when $\sigma_{1},\lambda_{2}$ are real. Note that the dispersion
equation (\ref{disp}) is written with this assumption in mind, since
all eigenvalues $\gamma$ must be real. First roots (ordering in increasing
order) of the dispersion equation in the regime when $\epsilon_{1}\gg\epsilon_{2}$
will appear in the intersection with the flat part of the hyperbolic
tangent curve, which can be very well approximated by a constant:
\[
\frac{\tanh(\pi\kappa_{2})}{\pi\kappa_{2}}\sim\frac{\tanh(\pi k_{y}/\sqrt{\epsilon_{2}})}{\pi k_{y}/\sqrt{\epsilon_{2}}},
\]
with $\tan(x)/x$ branches on intervals $(\pi/2+\pi n,3\pi/2+\pi n)$
for $n=1,2,\dots$. The tangent term can be approximated linearly
as 
\[
\frac{\tan(\pi\sigma_{1})}{\pi\sigma_{1}}\sim\frac{\sigma_{1}-n}{n},
\]
which leads to the eigenvalue estimates 
\begin{equation}
\gamma_{n}=\frac{k_{y}^{2}}{\epsilon_{1}}+n^{2}\left(1-\frac{\tanh(\pi k_{y}/\sqrt{\epsilon_{2}})}{\pi k_{y}/\sqrt{\epsilon_{2}}}\right)^{2}.
\end{equation}
Note that, for all parameters of interest, $\xi=\pi k_{y}/\sqrt{\epsilon}\gg1$
and $\tanh(\xi)/\xi\ll1$, thus eigenvalues grow quickly as $n^{2}$.
Therefore, the less damped and most important eigenvalue which controls
the time evolution most of the time (except perhaps during the initial
transient phase) is 
\begin{equation}
\gamma=\gamma_{1}=\frac{k_{y}^{2}}{\epsilon_{1}}+\left(1-\frac{\tanh(\pi k_{y}/\sqrt{\epsilon_{2}})}{\pi k_{y}/\sqrt{\epsilon_{2}}}\right)^{2}.\label{gamma1}
\end{equation}

\section{Performing the Fourier kernel integral}

\label{sec:kernel} In this section, we simplify the expression of
the Fourier kernel $\mathcal{K}(\mathbf{x})$ in (\ref{exact_step_kernel}).
First, we use the fact that 
\begin{align}
 & \int d^{2}k_{\perp}\frac{e^{i\mathbf{k}_{\perp}\cdot\mathbf{x}_{\perp}}}{1+ak_{\perp}^{2}}=2\pi\int_{0}^{\infty}dk_{\perp}\frac{k_{\perp}}{1+ak_{\perp}^{2}}\frac{1}{\pi}\int_{0}^{\pi}d\theta e^{ik_{\perp}r_{\perp}\cos\theta}=\\
= & 2\pi\int_{0}^{\infty}dk_{\perp}\frac{k_{\perp}}{1+ak_{\perp}^{2}}J_{0}(k_{\perp}r_{\perp})=\frac{2\pi}{a}K_{0}\left(\frac{r_{\perp}}{\sqrt{a}}\right),
\end{align}
where $a>0$ is a positive constant, $r_{\perp}=|\mathbf{x}_{\perp}|$,
$J_{0},K_{0}$ are the first-kind and modified second-kind Bessel
functions, respectively. The last step in the integration used equation
(11.4.44) on p488 in \citep{abramowitz1964handbook}. Now we apply
it to the kernel integral: 
\begin{equation}
\mathcal{K}(\mathbf{x})=\frac{1}{(2\pi)^{3}}\text{Re}\left[\int d^{3}ke^{i\mathbf{k}\cdot\mathbf{x}}\frac{e^{-\alpha}}{1+\beta(\frac{1-e^{-\alpha}}{\alpha})}\right],
\end{equation}
with $\beta=\Delta tk_{\perp}^{2}$ and get: 
\begin{align}
\mathcal{K}(\mathbf{x})=\frac{1}{4\pi^{2}\Delta t}\int_{-\infty}^{\infty}dk_{\parallel}\cos(k_{\parallel}z)e^{-\alpha}\frac{\alpha}{1-e^{-\alpha}}K_{0}\left(r_{\perp}\sqrt{\frac{\alpha}{1-e^{-\alpha}}}\right).
\end{align}

\bibliographystyle{model1-num-names}
\bibliography{ref}

\end{document}